\documentclass[12pt,preprint]{aastex}

\usepackage{natbib}
\usepackage[utf8]{inputenc}
\usepackage{graphicx}
\usepackage{amsmath}
\usepackage{lscape}

\begin{document}

\title{M Stars in the TW Hya Association: \\ Stellar X-rays and Disk Dissipation} 

\author{Joel H.\ Kastner\altaffilmark{1}, David A. Principe\altaffilmark{2,3}, Kristina Punzi\altaffilmark{1}, Beate Stelzer\altaffilmark{4}, Uma Gorti\altaffilmark{5}, Ilaria Pascucci\altaffilmark{6}, Costanza Argiroffi\altaffilmark{4,7}}

\altaffiltext{1}{Chester F. Carlson Center for Imaging Science, School of Physics \& Astronomy, and 
Laboratory for Multiwavelength Astrophysics, Rochester Institute of
Technology, 54 Lomb Memorial Drive, Rochester NY 14623 USA
(jhk@cis.rit.edu)}
\altaffiltext{2}{N\'{u}cleo de Astronom\'{i}a de la Facultad de Ingenier\'{i}a, Universidad Diego Portales, Chile}
\altaffiltext{3}{Millennium Nucleus Protoplanetary Disks, Chile}
\altaffiltext{4}{INAF-Osservatorio Astronomico di Palermo, Piazza del Parlamento 1, 90134, Palermo, Italy}
\altaffiltext{5}{SETI Institute, 189 Bernardo Ave., Mountain View, CA 94043, USA; NASA Ames Research Center, Moffett Field, CA 94035, USA}
\altaffiltext{6}{Lunar and Planetary Laboratory, The University of Arizona, Tucson, AZ 85721, USA}
\altaffiltext{7}{Dip.\ di Fisica e Chimica, Universit\`a di Palermo, Piazza del Parlamento 1, 90134, Palermo, Italy}

\begin{abstract}
To investigate the potential connection between the intense X-ray emission from young, low-mass stars and the lifetimes of their circumstellar, planet-forming disks, we have compiled the X-ray luminosities ($L_X$) of M stars in the $\sim$8 Myr-old TW Hya Association (TWA) for which X-ray data are presently available. Our investigation includes analysis of archival {\it Chandra} data for the TWA binary systems TWA 8, 9, and 13. Although our study suffers from poor statistics for stars later than M3, we find a trend of decreasing $L_X/L_{bol}$ with decreasing $T_{eff}$ for TWA M stars wherein the earliest-type (M0--M2) stars cluster near $\log{(L_X/L_{bol})} \approx -3.0$ and then $\log{(L_X/L_{bol})}$ decreases, and its distribution broadens, for types M4 and later. The fraction of TWA stars that display evidence for residual primordial disk material also sharply increases in this same (mid-M) spectral type regime. This apparent anticorrelation between the relative X-ray luminosities of low-mass TWA stars and the longevities of their circumstellar disks suggests that primordial disks orbiting early-type M stars in the TWA have dispersed rapidly as a consequence of their persistent large X-ray fluxes. Conversely, the disks orbiting the very lowest-mass pre-MS stars and pre-MS brown dwarfs in the Association may have survived because their X-ray luminosities and, hence, disk photoevaporation rates are very low to begin with, and then further decline relatively early in their pre-MS evolution.
\end{abstract}

\section{Introduction}

Thanks to their low luminosities and close-in habitable zones --- i.e., the range of exoplanet
orbital semimajor axes where water may exist in liquid form and, as a
result, life may eventually flourish --- the lowest-mass (M-type) stars
represent the best targets for future direct (imaging) giant planet searches and indirect (transiting) discovery and characterization of potentially habitable exoplanets. It is hence essential to establish, on both theoretical and observational grounds, whether planets are expected to be common around M stars. 
Based on the data available thus far, giant planets seem to be rare around M dwarfs, but terrestrial planets and super-Earths may be quite common \citep[e.g.,][]{Mulders2015a, Mulders2015b}. Indeed, the occurrence rate of 1--4 $R_{\rm Earth}$ planets around M dwarfs appears to be higher than that around solar-mass stars \citep{Howard2012,Mulders2015a,Mulders2015b}. 

Meanwhile, the answer to the corresponding, fundamental theoretical questions --- why should giant planets be rare and terrestrial planets common around M stars? --- requires understanding the harsh conditions out of which such planets form. Low-mass, pre-main sequence (pre-MS) stars are characterized by intense high-energy radiation fields \citep[e.g.,][]{Preibisch2005,Guedel2007}. This strong UV and X-ray emission has its origins in a combination of stellar magnetic and accretion activity. As low-mass pre-MS stars descend to the main sequence, their deep convective envelopes combine with differential rotation to produce strong magnetic dynamos and, hence, high levels of chromospheric and coronal activity; the former is a source of bright UV emission, while the latter generates strong X-ray emission \citep[e.g.,][and refs.\ therein]{Stelzer2013}.
In the case of T Tauri stars that are actively accreting from
circumstellar disks, shocks at the bases of accretion columns can also
significantly contribute to UV and soft X-ray emission \citep{Guenther2007,Sacco2010}.

The resulting irradiation of protoplanetary disks by
high-energy stellar photons likely regulates exoplanet formation and evolution processes. Various theoretical
studies have shown that EUV and X-ray radiation from young stars can drive
disk dissipation and disk chemistry, thereby determining the timescale
over which, and the conditions out of which, exoplanets and their
atmospheres emerge \citep[e.g.,][]{Gorti2009,Ercolano2009,Glassgold2012,Owen2012,Walsh2012,Cleeves2013,Gorti2015}. In particular, the EUV and soft ($\stackrel{<}{\sim}$1 keV) X-ray  radiation field of the central star should be a major source of disk surface heating and, as a result, potentially represents an important driver of slow, photoevaporative disk winds \citep{Gorti2009,Ercolano2009,Owen2012}.  There is observational evidence for the presence of such stellar EUV/X-ray-generated photoevaporative disk winds in the case of relatively evolved pre-main sequence stars of roughly solar mass \citep[][]{Pascucci2009,Sacco2012,ClarkeOwen2015}. Whether protoplanetary disks orbiting ultra-low-mass stars and brown dwarfs are similarly irradiated and actively photoevaporating remains to be determined \citep[see, e.g.,][]{Pascucci2013}.

To improve our understanding of the potential effects of X-rays on planet formation in disks orbiting M stars, we must characterize the X-ray emission properties of the lowest-mass (mid- to late-type M type) pre-MS stars of age $>$3 Myr, i.e., the epoch during or just after giant planet building and just preceding terrestrial planet building. The TW Hya Association \citep[TWA;][]{Kastner1997,Webb1999} affords just such an opportunity, thanks to its mean distance of just $\sim$50 pc and age $\sim$8 Myr \citep[][and refs.\ therein]{Torres2008,Ducourant2014,HerHill2015}.  Here, we present an analysis of all published and archival X-ray observations of M-type stars in the TWA in light of the presence or absence of evidence for gas and dust in circumstellar disks around these same stars, so as to further constrain the potential relationship between X-rays and disk dispersal.

\section{TW Hya Association M stars: X-ray luminosities and disk detection rates}

\begin{table*}[!htbp]
\footnotesize
\begin{center}
\caption{\sc M-type TWA Members with Available X-ray Data}
\label{tbl:GenData}
\vspace{.1in}
\begin{tabular}{lccccc}
\hline
\hline
ID & sp.\ type$^a$ & $D$$^b$ & $J$ & $L_{bol}$$^c$ & mid-IR excess$^d$ \\ 
     &                & (pc)  & (mag) & ($\times 10^{32}$ erg s$^{-1}$) &  \\
\hline
TWA 6   &        M0    &     51 &         8.87 &    4.72 & N \\
TWA 25 &	M0.5 &	51 &	8.17 &    9.04 & N \\
TWA 14 &	M0.5 &	113 &	9.42 &    13.8 & N \\
TWA 13A &	M1 &        55 &        8.43 &     7.86 & N \\
TWA 13B &	M1 &	55 &	8.43 &	7.88 & N \\
TWA 16 &	M1 &	65 &	8.99 &	6.54 & N \\
TWA 21 &	M1 &	50 &	7.87 &	11.1 & N \\
TWA 23 &	M1 &	48 &	8.62 &	5.12 & N \\
TWA 5A &	M1.5+M2 & 49 &	7.67	&	11.8 & N \\
TWA 2AB &	M2+M2 & 42 &	7.63 &	8.91 & N \\
TWA 10 &	M2 &	62 &	9.12 &	4.92 & N \\
TWA 11B &	M2 &	67 &	9.15 &	5.68 & ... \\
TWA 12 &	M2 &	65 &	9.00 &	6.18 & N \\
TWA 7 &	M3 &	35 &	7.79 &	        5.08 & Y \\
TWA 8A &	M3 &	43 &	8.34 &	4.73 & N \\
TWA 9B &	M3 &	52 &	9.98 &	1.54 & N \\
TWA 15B &	M3 &	113 &	10.49 &	4.56 & N \\
TWA 20 &	M3 &	73 &	9.33 &	5.51 & N \\
TWA 3A &	M3+M4 &	31 &	7.65 & 2.70 & Y \\
TWA 15A &	M3.5 &	113 &	10.56 &	4.25 & N \\
TWA 3B &	M4 &	31 & ... &  1.68 & N \\	
TWA 30B &	M4 &	44 & 15.35 & 2.0 & Y \\	
TWA 31 &	M4.2 &	110	& 13.05 & 0.4 & Y \\	
TWA 11C &	M4.5 &	67 &	9.79	 &    2.75 & N \\
TWA 8B &	M5 &	39 &	9.84 &	0.83 & N \\
TWA 30A &	M5 &	42 &	9.64 &	1.2 & Y \\
TWA 5B &	M8 &	49 &	12.60 & 0.08 & ... \\	
TWA 27 &	M8 &	53 &	13.00 &0.12 & Y \\	
TWA 28 &	M8.5 &	55 &	13.03 &0.075 & Y \\	
TWA 26 &	M9 &	38 &	12.69 &0.09 & N \\	
\hline
\end{tabular}
\end{center}

\footnotesize
a) Spectral types from \citet{Schneider2012}, \citet{Manara2013}, and \citet{HerHill2014};
see text. b) Trigonometric parallax distances from
\citet{Ducourant2014}  where available; otherwise, distances from
\citet{Schneider2012}. c) Bolometric luminosities obtained from $J$
and spectral-type-dependent bolometric corrections listed in
\citet{PecautMamajek2013}, except for TWA 30A and 30B ($L_{bol}$ from
Principe et al.\ 2015 and refs.\ therein), TWA 5B \citep[$L_{bol}$
from][]{Tsuboi2003}, TWA 27 \citep[$L_{bol}$ from][]{GizisBharat2004},
TWA 28 \citep[$L_{bol}$ from][]{Scholz2005}, and TWA 26 \citep[$L_{bol}$
from][]{Castro2011}. d) Mid-IR excess (or lack thereof) as determined from
WISE data by \citet{Schneider2012} or in this study (see text). The
presence of absence of mid-IR excess is unknown for TWA 5B
and 11B due to proximity to the (far more mid-IR-luminous) primary stars
in these systems.
\end{table*}

The sample considered here (Table~\ref{tbl:GenData}) consists of those high-probability M-type members of the TWA for which archival or published X-ray data from ROSAT (All-Sky Survey; RASS), Chandra, or XMM-Newton were available at the time this paper was written, based on a search of the HEASARC \verb+browse+ search utility\footnote{https://heasarc.gsfc.nasa.gov/cgi-bin/W3Browse/w3browse.pl} and the literature. Appendix A contains notes concerning the sample's inclusion and exclusion of specific TWA members and candidate members. For each sample star, we list in Table~\ref{tbl:GenData} the spectral type and distance we have adopted from the literature, as well as 2MASS $J$ magnitude and bolometric luminosity $L_{bol}$.  For known binaries that are unresolved in the available X-ray data (TWA 2AB, 3A, and 5A), we list composite spectral types and total $L_{bol}$ values. The listed spectral types are obtained from \citet[][and refs.\ therein]{Schneider2012}, \citet{Manara2013}, and \citet{HerHill2014}. The spectral types of TWA 14 (M0.5), TWA 13AB (both M1), TWA 2A (M2), TWA 8A (M3) and TWA 8B (M5) listed in \citet{Schneider2012} were confirmed in one or both of the \citet{Manara2013} and \citet{HerHill2014} studies, while one or both of these studies determined later spectral types for TWA 7 (M3), TWA 9B (M3), TWA 15A (M3.5), TWA 15B (M3), and TWA 25 (M0.5). We adopt these revised classifications here.
Except where noted, the $L_{bol}$ values were obtained from the listed spectral types, stellar distances, and $J$ band data based on the (spectral-type-dependent) $J$ band bolometric corrections determined by \citet{PecautMamajek2013}, assuming no reddening. 

\subsection{X-ray luminosities}

\begin{table}[!htbp]
\footnotesize
\begin{center}
\caption{\sc M-type TWA Members: X-ray Luminosities}
\label{tbl:XrayData}
\vspace{.1in}
\begin{tabular}{lccccccc}
\hline
\hline
 ID & sp.\ type  & \multicolumn{4}{c}{$L_X$, 0.3--8.0 keV ($10^{29}$
                   erg s$^{-1}$)} & $\log{L_X/L_{bol}}$ & Notes  \\
 & & ROSAT & XMM & Chandra & adopted \\
(1) & (2) & (3) & (4) & (5) & (6) & (7) & (8) \\
\hline
TWA 6      & M0    & 5.4 & ... & 10.3 &  10.3 & $-2.66$ \\
TWA 25	& M0.5 & 11.3 & ... & ... &				11.3	& $-2.90$ \\
TWA 14	& M0.5 &	10.8 & ... & ... &			10.8	& $-3.11$ \\
TWA 13A	& M1 &	(2.3) &	(5.1) &	10.7 &	9.2 & $-2.93$ & 1\\
TWA 13B	& M1 &	(2.3) &	(5.1) &     12.9 &	11.1	& $-2.85$ & 1\\
TWA 16	& M1 &	3.2 & ... & ... &                        3.2 &	$-3.32$ \\
TWA 21	& M1 &	5.9 & 	6.1 & ... & 		6.1 &	$-3.23$ \\
TWA 23	& M1 &    2.5 & ... & ... &		        2.5	& $-3.31$ \\
TWA 5A	& M1.5+M2 &	12.0 & 7.0 & 	9.4 &	6.5	& $-3.26$ & 2 \\
TWA 2AB	& M2+M2 &	4.5 & ... & ... &	        2.1	& $-3.62$ & 3 \\
TWA 10	& M2 &	3.2 &		6.0 & ... &	6.0	&  $-2.91$ \\
TWA 11B	& M2 & ... & ... & ... &				4.5	& $-3.10$ & 4 \\
TWA 12	& M2 &	3.5 & 8.5 & ... &		        8.5	& $-2.86$ \\
TWA 7	& M3 &	2.7 & 3.3 & 7.2 & 	                4.7	& $-3.03$ & 5 \\
TWA 8A	& M3 &	(1.9) & ... & ... &			6.3	& $-2.88$ & 1 \\
TWA 9B	& M3 &	(3.3) & ... & ... &			0.64	& $-3.38$ & 1 \\
TWA 20	& M3 &	2.7 & ... & ... & 			2.7	& $-3.31$ \\
TWA 15AB & M3+M3.5 &	1.3 & ... & ... &		1.3	& $-3.88$ \\
TWA 3A	& M3+M4 & (1.0) & ... & ... &                       1.0 & $-3.43$ & 6\\
TWA 3B	& M4 & (1.0) & ... & ... &                       2.0 & $-2.92$& 6 \\		
TWA 30B	& M4 & ... & ... & ... &				$<$0.030	& $<$-4.8 & 7 \\
TWA 31	& M4.2 & ... & $<$0.034 & ... &				$<$0.034	& $<$-4.1 \\
TWA 11C	& M4.5 & ... & ... & ... &				1.2	& $-3.37$ & 8 \\
TWA 8B	& M5 &(1.9) & ... & ... &			        0.46	& $-3.25$ & 1 \\
TWA 30A	& M5 & 1.4 & ... & ... &			0.12 &	$-4.00$ & 7 \\
TWA 5B	& M8 & ... & ... & ... &				0.03 & $-3.40$ & 9 \\
TWA 27	& M8 & ... & ... & ... &                                $<$0.0012 & $<$-5.1 & 10\\
TWA 28	& M8.5 & ... & ... & ... &				$<$0.0053 & $<$-4.1 & 11\\
TWA 26	& M9 & ... & ... & ... &				0.0014	& $-4.80$ & 12 \\
\hline
\end{tabular}
\end{center}

\footnotesize
1) Binary unresolved by ROSAT and XMM. Adopted component $L_X$ values
determined from the analysis of Chandra data described in Appendix B. 2) Adopted $L_X$
determined from Chandra data \citep{Tsuboi2003}. 3) Adopted $L_X$
determined from Swift data \citep{Brown2015}. 4) Adopted $L_X$
determined from Chandra serendipitous observation (D. Huenemoerder,
pvt.\ comm.). 5) Adopted $L_X$ determined from Chandra data
\citep{Brown2015}. 6) Binary unresolved by ROSAT; adopted component
$L_X$ values determined from Chandra data
\citep{Huenemoerder2007}. 7) Adopted $L_X$ determined from XMM data
(Principe et al.\ 2015, and refs.\ therein).  8) Adopted $L_X$
determined from XMM data \citep{Kastner2008}. 9) Adopted $L_X$
determined from Chandra data \citep{Tsuboi2003}. 10) Adopted $L_X$
determined from Chandra data \citep{GizisBharat2004}. 11) Adopted
limit on $L_X$ determined from XMM-Newton data
\citep{Stelzer2007}. 12) Adopted $L_X$ determined from Chandra data
\citep{Castro2011}. 
\end{table}

Table~\ref{tbl:XrayData} lists X-ray luminosities ($L_X$) for the sample stars. The values of $L_X$ in columns 3, 4 and 5 of Table~\ref{tbl:XrayData} are all calculated over the energy range 0.3--8.0 keV from available archival (HEASARC database) ROSAT All-sky Survey (RASS), XMM-Newton, and Chandra count rates, respectively, as described below. Note that whereas X-ray count rates (hence $L_X$ values) are available for all but one of the early-M (M0 to M2) TWA members from the RASS, only a few TWA stars of type M3 or later were detected in the RASS, as a consequence of its limited sensitivity. We hence do not consider RASS count rate upper limits in the present study, as these upper limits do not place meaningful constraints on $L_X$ for young mid- to late-M stars at the distance of the TWA \citep{Rodriguez2013}.  

To obtain the values of $L_X$ in columns 3--5, we converted the count rates to 0.3--8.0 keV X-ray fluxes ($F_X$) via the \verb+webpimms+ tool\footnote{https://heasarc.gsfc.nasa.gov/cgi-bin/Tools/w3pimms/w3pimms.pl}, which accounts for the different energy sensitivities of the three missions (the Chandra and XMM-Newton X-ray Observatories and their back-illuminated CCD sensors cover the entire 0.3--8.0 keV range, whereas the Position Sensitive Proportional Counter aboard ROSAT was sensitive in the range $\sim$0.1-2.0 keV). We assumed a single-component absorbed thermal plasma model whose parameters are characteristic plasma temperature ($T_X$), metallicity relative to solar, and intervening absorbing column ($N_H$). We adopted parameter values of $kT_X = 1.0$ keV ($T_X \approx 12$ MK), $N_H = 10^{19}$ cm$^{-2}$, and a metallicity of 0.2. These choices for parameter values are based on the results of model fitting to the XMM-Newton spectra of TWA~11B \citep[][and refs.\ therein]{Kastner2008} and TWA~30A \citep{Principe2015}, as well as the results of Chandra X-ray spectral fitting presented in Appendix B for the individual components of the binary systems TWA 8AB, 9AB, and 13AB. For these model parameter values, the count rate to $F_X$ conversion factors are $6.3\times10^{-12}$, $3.2\times10^{-12}$, and $4.1\times10^{-12}$ erg cm$^{-2}$ count$^{-1}$ for ROSAT, XMM-Newton, and Chandra, respectively. For binaries that are unresolved by ROSAT (TWA 3AB, 8AB, 9AB, and 13AB), we have split the ROSAT-based $F_X$ (column 3) equally between components. We then calculated $L_X$ from the values of $F_X$, assuming the distances listed in Table~\ref{tbl:GenData}. 

It is evident from Table~\ref{tbl:XrayData} that all of the stars for which data is available from multiple observatories (i.e., multiple epochs) show variability at the  level of a factor $\sim$2--3, as is typical for low-mass pre-MS stars \citep[e.g.,][]{Guedel2007,Principe2014}. The most extreme case in Table~\ref{tbl:XrayData} is the M5 star TWA 30A, which displayed at least a factor $\sim$10 change (decrease) in $L_X$ between the (1990) RASS and (2011) XMM-Newton observing epochs \citep{Principe2015}.

In columns 6 and 7 of the Table we list the values of $L_X$ and $\log{(L_X/L_{bol})}$, respectively, that we have adopted for the analysis described in \S 3. Wherever possible, these adopted values of $L_X$ were obtained from analyses of Chandra or XMM-Newton spectra. Specifically, for the binary systems TWA 8, 9, and 13, the $L_X$ values were determined from our analysis of archival Chandra data (Appendix B), while for the stars TWA 5A, 5B, 7, 11C, 26, 27, 28, 30A, and 30B, the adopted $L_X$ values in column 6 were obtained from the literature (see references listed in column 8 of Table~\ref{tbl:XrayData}). In adopting a value of $L_X$ for each star, Chandra results have been given priority, since Chandra most reliably resolves the X-ray counterparts to TWA binary systems. In cases where both XMM-Newton and RASS observations (but no Chandra data) are available, we adopt the XMM-based $L_X$. Comparing the values of $L_X$ calculated for TWA 5A, 7, 13A and 13B based on the single-component \verb+webpimms+ model (column 5) with values obtained from spectral modeling of the same (Chandra) data (column 6), it appears that our use of the \verb+webpimms+ model may systematically overestimate $F_X$ by $\sim$30\%. This potential level of systematic error does not affect the results described in \S 3.

\subsection{Presence or absence of circumstellar disks}

In the last column of Table~\ref{tbl:GenData} we indicate whether each sample star displays a mid-infrared excess indicative of the presence of warm ($T \sim$ 100--300 K) dust in a circumstellar disk. These assessments of the presence of absence of dusty disks are drawn from the analysis of WISE data by \citet{Schneider2012}, with the exception of TWA 14, 15AB, and 21 (which were omitted from their sample; see Appendix A). The WISE point source catalog $W1-W4$ colors of these three stars are 0.25, 0.20, and 0.08, respectively (where the TWA 15AB binary is unresolved by WISE), all of which lie well within the locus of TWA star colors for stars that lack IR excesses \citep{Schneider2012}.  The mid-IR excess stars TWA 3A, 7, 30A, 30B, and 31 also display relatively strong far-IR fluxes \citep{Riviere2013,Liu2015}. Reanalysis of archival HST imaging has yielded a direct detection of the TWA 7 dust disk via scattered starlight; TWA 25 also displays a compact, nearly edge-on dust disk in HST imaging despite its lack of detectable mid-IR excess \citep{Choquet2015}.

Of the seven Table~\ref{tbl:GenData} stars that display mid-IR excesses, all but one \citep[TWA 7;][]{Manara2013} also show evidence for active accretion of gas from their disks in the form of unusually strong, broad H$\alpha$ emission \citep[see \S 3.2 and][]{Muzerolle2000,Stelzer2007,Looper2010,Shkolnik2011}. The variable, weak accretors TWA 27 and 28 also display He~{\sc i} line emission \citep{Herczeg2009}. In the case of the $\sim$1.4$''$ separation binary TWA 3AB, Chandra X-ray spectroscopy (and large UV excess) also indicate the presence of accretion shocks in this system, with the X-ray signature of accretion more evident at the dusty binary component \citep[TWA 3A;][]{Huenemoerder2007}. But the most extreme examples of gas-rich disks among the TWA M stars with mid-IR excesses are the two components of the wide binary TWA 30A and 30B. These star/disk systems, both of which are evidently viewed nearly edge-on, display evidence for the presence of circumstellar gas in the form of forbidden emission lines detected via optical spectroscopy \citep{Looper2010} and, in the case of TWA 30A, in the form of attenuation of stellar X-rays \citep{Principe2015}. The relatively weak H$\alpha$ emission seen toward TWA 30A and 30B (emission-line equivalent widths $<10$\AA) can be ascribed to their viewing geometry, which in each case at least partially obscures the regions of active accretion onto the star \citep{Looper2010,Looper2010a}. Furthermore, during a photometric monitoring campaign, TWA 30A was seen to display quasi-periodic dips in its multi-band optical and near-IR lightcurves, bolstering the \citet{Looper2010a} model invoking disk structures that rotate in and out of the line of sight, temporarily obscuring the star \citep{Stelzer2015}.

Knowledge of the presence or absence of cold ($T < 100$ K) circumstellar dust and gas is scant for the Table~\ref{tbl:GenData} stars. Only three of these stars --- TWA 30A, 30B, and 31 --- have been the subject of sensitive mm- and submm-wave continuum and CO observations \citep[with the Atacama Large Millimeter Array;][]{Rodriguez2015}. All three display mid- to far-IR excesses indicative of warm dust \citep{Schneider2012,Liu2015}. However, only TWA 30B was detected as a submm continuum source, and none of the three were detected as CO sources, in the \citet{Rodriguez2015} ALMA survey of ultra-low-mass members and candidate members of the TWA. The CO nondetections imply the gas disks orbiting these three stars have very low masses and/or small radii \citep[$\stackrel{<}{\sim}$0.1 Earth masses and/or $\stackrel{<}{\sim}$10 AU, respectively;][]{Rodriguez2015}.

\subsection{Trends in X-ray luminosity and disk fraction}

\begin{figure*}[!h]
\centering
\includegraphics[height=3.in,angle=0]{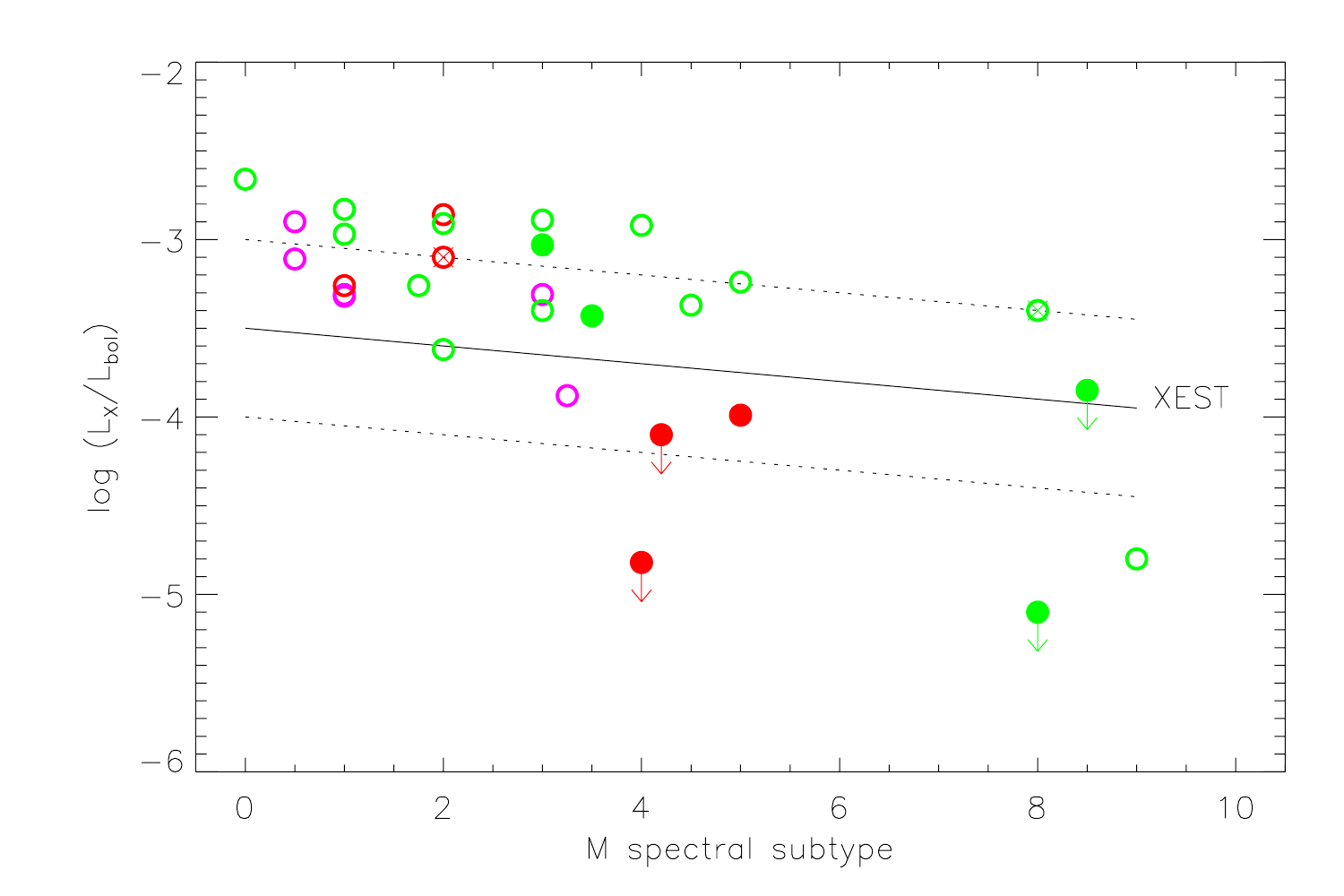}
\includegraphics[height=3.in,angle=0]{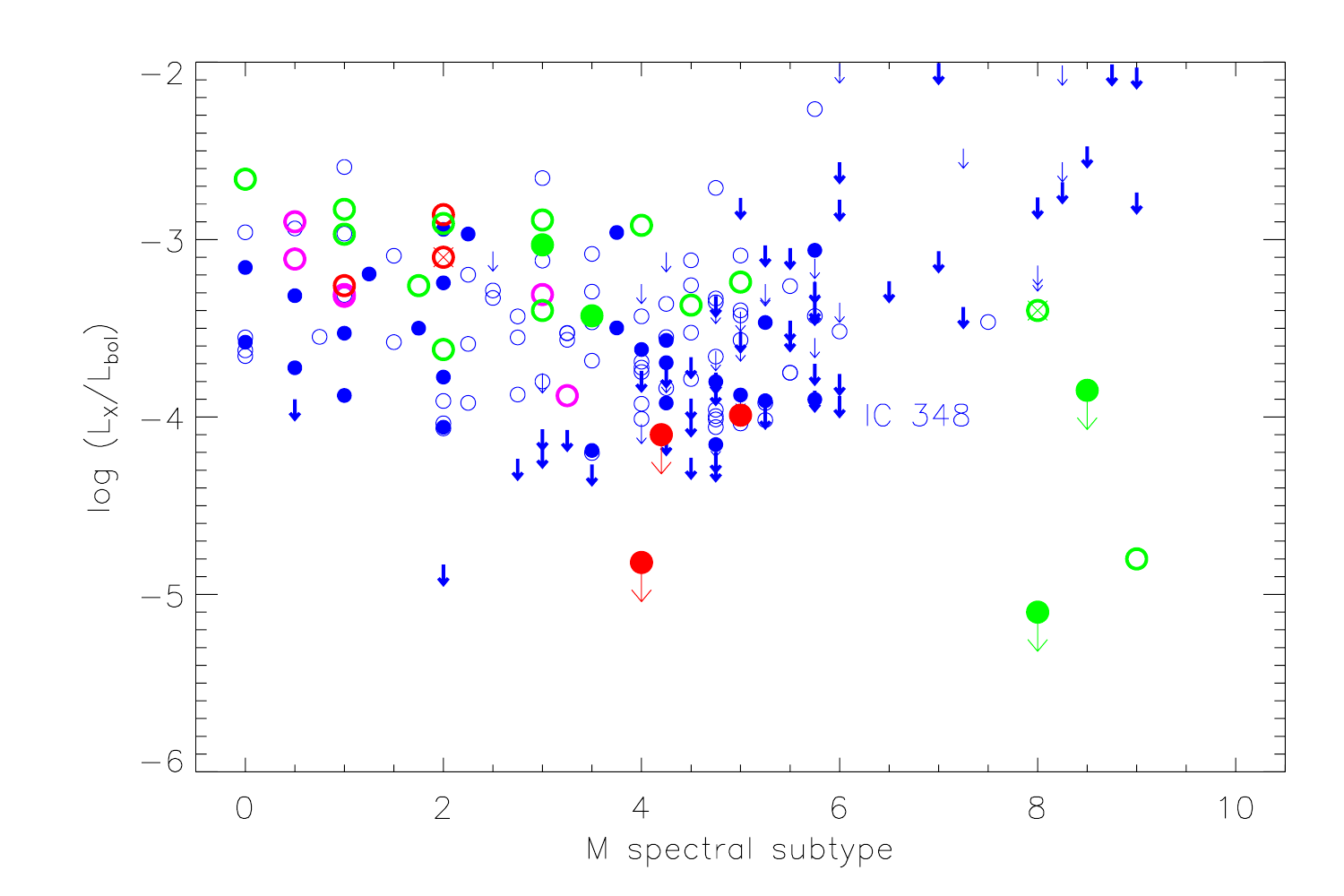}
\vspace{-.18in}
\caption{\footnotesize 
{\it Top:} Log of the ratio of X-ray to bolometric luminosity for TWA M-type stars, plotted as a function of spectral subclass. Magenta, red, and green symbols indicate ROSAT, XMM, and Chandra observations, respectively. Filled and open circles represent stars with and without warm circumstellar dust, i.e., with and without detectable mid-IR excesses; an `X' indicates the IR excess status of the star is unknown (see Table~\ref{tbl:GenData}). The solid line indicates the dependence of $\log{(L_X/L_{bol})}$ on spectral type, and dashed lines the 1 $\sigma$ scatter in $\log{(L_X/L_{bol})}$, as determined by \citet{Grosso2007} for low-mass T Tauri stars and young brown dwarf candidates that were included in the XMM-Newton extended survey of Taurus \citep[XEST;][]{Guedel2007}. {\it Bottom:} As in the top panel, with blue symbols indicating measurements of $\log{(L_X/L_{bol})}$ for low-luminosity stars in IC 348 \citep{Stelzer2012}, where filled and open circles represent X-ray-detected pre-MS stars with disks (Class II) and without disks (Class III), respectively (and thick and thin arrows indicate X-ray-nondetected stars of these respective classes).}
\label{fig:lxlbolsptype}
\end{figure*}

\begin{figure*}[!h]
\centering
\includegraphics[width=5in,angle=0]{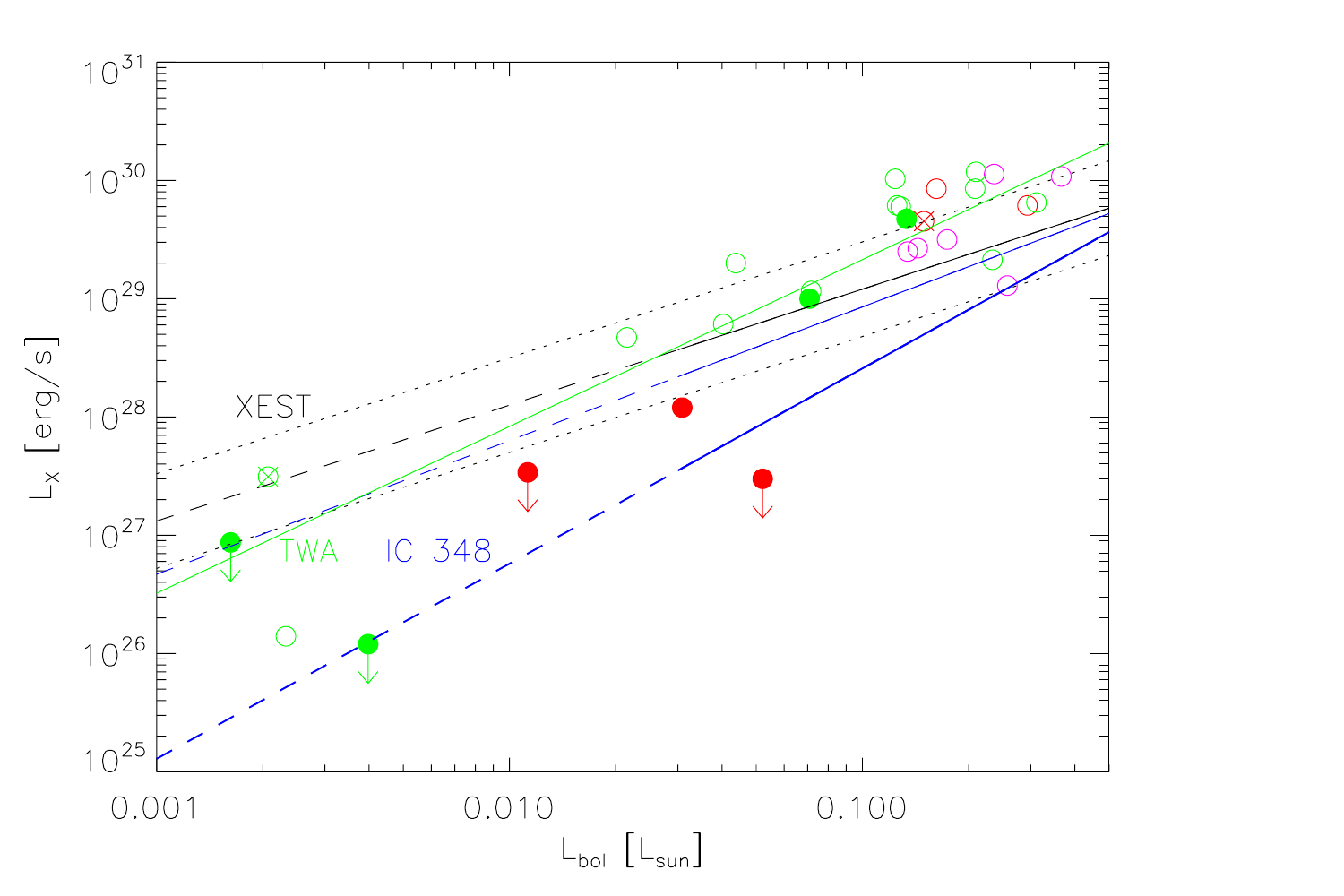}
\vspace{-.18in}
\caption{\footnotesize 
X-ray vs.\ bolometric luminosities for TWA M type stars, with symbols as in Fig.~1. The best-fit linear regression is indicated as a green line. The empirical relationships between $L_X$ and $L_{bol}$ inferred for low-mass T Tauri stars and young brown dwarf candidates in Taurus \citep[][]{Grosso2007} and for low-luminosity stars in IC 348 \citep{Stelzer2012} are indicated by the black and (two) blue lines, respectively; the two lines for IC 348 are the results of linear regression results for Class II and Class III pre-MS stars (lower and upper blue lines, respectively). The dashed portions of the black and blue lines indicate extrapolations of the linear regressions of $L_X$ vs.\ $L_{bol}$ for Taurus and IC 348, and the dotted black lines indicate the approximate 1$\sigma$ scatter for the Taurus sample. }
\label{fig:lxlbol}
\end{figure*}

In Fig.~\ref{fig:lxlbolsptype}, we plot the log of the ratio $L_X/L_{bol}$ as a function of M spectral subtype for the TWA M stars. A trend is apparent, in which the earliest-type (M0--M2) stars cluster near $\log{(L_X/L_{bol})} \approx -3.0$, and then $\log{(L_X/L_{bol})}$ decreases, and its distribution broadens, for spectral types M4 and later. As Fig.~\ref{fig:lxlbolsptype} demonstrates, such a trend of decreasing $\log{(L_X/L_{bol})}$ with M subtype is also apparent for the younger pre-MS M stars in the Taurus molecular clouds \citep[median age $\sim$0.5 Myr;][]{Grosso2007} and for M stars in the young cluster IC 348 \citep[age $\sim$3 Myr;][]{Stelzer2012}. The decline of $\log{(L_X/L_{bol})}$ with decreasing M star effective temperature appears to be somewhat steeper in the TWA than in Taurus, and is similar to IC 348 in the range M0--M5. The $L_X/L_{bol}$ ratios of early-M TWA stars appear to be systematically larger than those of their counterparts in Taurus and IC 348 (but see below). 

In Fig.~\ref{fig:lxlbol}, we plot $L_X$ vs.\ $L_{bol}$ for the TWA M stars, and overlay the empirical relationships determined for pre-MS low-mass stars and brown dwarf candidates in Taurus \citep{Grosso2007} and IC 348 \citep{Stelzer2012}. As was also clear in Fig.~\ref{fig:lxlbolsptype} (top panel), it seems that the more luminous early-M type TWA stars (i.e., those with $L_{bol} > 0.1$ $L_\odot$) are overluminous in X-rays, relative to Taurus TTS in this same range of bolometric luminosity. The mid- to late-M type (lowest-luminosity) TWA stars (those with $L_{bol} < 0.02$ $L_\odot$) also appear underluminous in X-rays relative to their Taurus counterparts, although the comparison in this $L_{bol}$ range suffers both from the small number statistics of our TWA sample and a lack of X-ray detections for the Taurus sample \citep{Grosso2007}. 

The latter caveat notwithstanding, we find the slope of $\log{L_X}$ vs.\ $\log{L_{bol}}$ for the X-ray-detected TWA M stars, $1.41 \pm 0.14$, is significantly steeper than that determined for the $\sim$0.5 Myr-old M stars in Taurus \citep[0.98$\pm$0.06 for X-ray-detected stars;][]{Grosso2007}, and that the zero-$\log{(L_{bol}/L_\odot)}$ intercept for the TWA sample, $\log{L_{X,0}} = 30.73$, is larger than found for Taurus \citep[$\log{L_{X,0}} = 30.06$;][]{Grosso2007}. The latter comparison may reflect the fact that there are no known examples of actively accreting stars in the TWA in the M0--M2 spectral type range (see below and \S 3.1). Indeed, the X-ray luminosities of early-type M stars in the TWA are similar to those of Class III (diskless) early-M stars in IC 348  (Fig.~\ref{fig:lxlbolsptype}, bottom panel). As illustrated in Fig.~\ref{fig:lxlbol}, the slope of $\log{L_X}$ vs.\ $\log{L_{bol}}$ we find for the TWA --- which includes stars with and without disks, and spans the range $L_{bol} =$ 0.015...0.3 $L_\odot$ --- is intermediate between the slopes determined for Class II and Class III pre-MS stars in IC 348 in the range $L_{bol} =$ 0.03...3.0 $L_\odot$, i.e., 1.13$\pm$0.11 and 1.65$\pm$0.22, respectively \citep[where these determinations take into account upper limits on $L_X$;][]{Stelzer2012}. 

The overall disk fraction among M-type TWA members with available X-ray data (Table~\ref{tbl:GenData}) is $\sim$25\% (7 of 28, excluding the two stars for which the presence or absence of disks cannot be established due to the proximity of IR-bright companions). This disk fraction is similar to but somewhat smaller than that inferred for TWA members overall \citep{Schneider2012}. However, Figs.~\ref{fig:lxlbolsptype} and \ref{fig:lxlbol} also make apparent that the fraction of TWA M stars with evidence for circumstellar disks increases as stellar mass decreases. Specifically, 6 of 11 stars with spectral types M3.5 and later ($\sim$50\%) have detectable warm circumstellar dust. All 6 of these stars also display evidence for circumstellar gas (\S 2.2). In contrast, the disk fraction among stars of type M3 and earlier drops to $\sim$5\%, i.e., one in 17, with the lone exception being the debris disk orbiting the M3 star TWA~7. The apparent jump in disk fraction hence occurs very near the same (M3/M4) spectral type boundary where $L_X/L_{bol}$ appears to decline (Fig.~\ref{fig:lxlbolsptype}). \citet{Schneider2012} noted a similar trend in which candidate substellar members of the TWA are more likely to display evidence for disks than are stellar members.

\section{Discussion}

\subsection{Implications for the early evolution of X-ray emission in M stars}

The comparison of the $\log{(L_X/L_{bol})}$ and $L_X$ distributions of the TWA M stars with the corresponding distributions for the much younger pre-MS M stars in Taurus (Figs.~\ref{fig:lxlbolsptype}, \ref{fig:lxlbol}) suggests that over the first $\sim$8 Myr of the lifetime of an early-M pre-MS star --- an epoch during which $L_{bol}$ is monotonically decreasing --- coronal X-ray luminosity remains roughly constant or perhaps even increases. 
In contrast, based on the sparse data presently available for mid- to late-M stars in the TWA, it appears that X-ray luminosity has decreased at least as fast as --- and, in at least some cases, much faster than --- bolometric luminosity for these stars (Fig.~\ref{fig:lxlbolsptype}). It is possible that gas in the disks orbiting the mid- to late-type M type TWA stars is attenuating their X-ray emission. However, such an explanation for the smaller values of $\log{(L_X/L_{bol})}$ observed for these stars appears unlikely, since the majority of the TWA M stars with disks do not show evidence for extinction of their photospheres by disk dust --- the notable exceptions being TWA 30A and 30B \citep{Looper2010,Looper2010a,Principe2015,Stelzer2015}.

The X-ray data presently available for the TWA, in combination with similar studies for Orion, Taurus, and IC 348 \citep{PreibischEtal2005,Grosso2007,Stelzer2012}, thereby potentially shed further light on the age at which coronal activity declines for stellar masses near the main sequence H-burning limit of
$\sim$0.08 $M_\odot$. Such a decline is apparent when comparing the X-ray
luminosities of very young (age $\stackrel{<}{\sim}$1 Myr) ultra-low-mass stars and brown dwarfs with those of (old) late-M and L-type field stars. Specifically, X-ray-detected pre-MS brown dwarf candidates in Orion and Taurus have typical X-ray luminosities $L_X \sim 10^{28}$ erg s$^{-1}$ \citep[][]{PreibischEtal2005,Grosso2007}, whereas $L_X \stackrel{<}{\sim} 10^{26}$ erg s$^{-1}$ for ultra-low-mass stars and brown dwarfs in the field \citep[e.g.,][and refs.\ therein]{Stelzer2006,Williams2014}. 

We note that the spectral type boundary where the ($\sim$8 Myr-old) TWA stars appear to
display a decline in $L_X/L_{bol}$ --- i.e., near M4 --- approximately corresponds to the dividing line  between pre-MS stars that will and will not undergo core H burning once on the main sequence, according to evolutionary models \citep[e.g.,][]{DAntona1997}. However, this is likely just a coincidence. In very low mass field stars (with ages of order Gyr), a drop in magnetic activity is observed around M8 \citep[e.g.,][]{West2004,Berger2006,Stelzer2006,Berger2010}, i.e., at an effective temperature well above the H-burning limit. The likely cause of this decrease of X-ray (and H$\alpha$) activity is the low level of ionization within the cool atmospheres of the latest-type M stars \citep[e.g.,][]{Mohanty2002}. 

The same (low ionization) effect also might play a role in the coolest (i.e., M8--M9) objects among our TWA sample, but would not explain the apparent low X-ray activity levels of three of the six mid-M stars (TWA 30A, 30B, and 31; Table~\ref{tbl:XrayData}). In that regard, it is interesting that all three of these stars are actively accreting from their dusty disks (\S\ 2.3). The apparent bifurcation of $L_X/L_{bol}$ in this spectral type range that is hinted at by Fig.~\ref{fig:lxlbolsptype} may therefore reflect the suppression of coronal X-ray emission by accretion. 
Alternatively, the internal (e.g., convective and/or rotational) structures of young, ultra-low-mass stars may differ fundamentally from those of higher-mass young stars \citep[see discussion in][and refs.\ therein]{Stelzer2012}. X-ray observations of additional mid- to late-type M stars of age $\sim$10 Myr, combined with spectroscopic diagnostics of mass accretion rates, will be necessary to distinguish between these different scenarios for the apparent overall drop in $L_X/L_{bol}$ for such stars at this age.

\subsection{Implications for disk dispersal via photoevaporation}

It is difficult to establish whether the dust responsible for the mid-IR excesses associated with M stars in the TWA resides in primordial (gas-rich) disks or (gas-poor) debris disks. Due to their small radial extent 
and low masses, the gaseous constituents of these disks are exceedingly difficult to detect in (sub)mm-wave molecular line emission. Indeed, as noted in \S 2.2, gas has been detected in the disks orbiting TWA 30A and 30B in the form of optical forbidden-line emission from disk winds and jets \citep{Looper2010a,Looper2010} as well as X-ray absorption by the (nearly edge-on) TWA 30A disk itself \citep{Principe2015}; however, a sensitive search for CO in these same disks with ALMA yielded negative results \citep{Rodriguez2015}. 

Nonetheless, there is reason to conclude that almost all of the TWA M stars with mid-IR excesses due to warm dust in fact host primordial disks that, given the $\sim$8 Myr age of the TWA, are highly evolved and in the process of dispersing. Specifically, a majority of these stars display H$\alpha$ emission indicative of active accretion of disk gas (Table~\ref{tbl:accretion}) and [O {\sc i}] $\lambda$6300 emission indicative of residual disk gas that is either orbiting or flowing out in a disk wind or jets. The components of the TWA 30 binary represent extreme cases in terms of [O {\sc i}] equivalent width \citep{Looper2010}, but [O {\sc i}]  emission has also been detected from TWA 3A, 27, and 28 \citep{Herczeg2009}. Among the TWA M stars with mid-IR excesses, only TWA 7 appears to harbor a gas-poor debris disk \citep{Riviere2013,Choquet2015}.

If most of the disks orbiting M stars in the TWA are indeed primordial, then the anticorrelation between X-ray luminosity and disk frequency noted in \S 2.3 (and apparent in Figs.~\ref{fig:lxlbolsptype}, \ref{fig:lxlbol}) may have implications for models describing how high-energy photons from low-mass stars can play a central role in driving disk dispersal via photoevaporation \citep[][and refs.\ therein]{Gorti2009,Ercolano2009,Ercolano2010,Owen2012,Gorti2015}. Specifically, models formulated by \citet[][and refs.\ therein]{Owen2012} predict that the rate of stellar-X-ray-induced disk gas
photoevaporation $\dot{M}_X$ is directly proportional to the stellar X-ray luminosity $L_X$,
\begin{equation}
\dot{M}_X = 8\times10^{-9} L_{X30} \: M_\odot \; {\rm yr}^{-1},
\end{equation}
where $L_{X30}$ is X-ray luminosity integrated over the energy range 0.1--10 keV in units of $10^{30}$ erg s$^{-1}$ assuming a relatively hard stellar X-ray spectral energy distribution (see below).

Eq.\ 1 predicts that, given their present X-ray luminosities  ($L_{X30} \approx 1.0$; Fig.~\ref{fig:lxlbol}), $\dot{M}_X\approx10^{-8}$ $M_\odot$ yr$^{-1}$ for early-M stars in the TWA. The predicted present-day photoevaporative mass loss rates for stars of type M4 and later, for which $L_{X30} \stackrel{<}{\sim} 0.1$, are at least an order of magnitude smaller. Viewed in this context, the results in Figs.~\ref{fig:lxlbolsptype}, \ref{fig:lxlbol} could be taken to suggest that primordial disks orbiting early-type M stars have dispersed rapidly as a consequence of their persistent large X-ray fluxes, while disks orbiting the very lowest-mass pre-MS stars and pre-MS brown dwarfs can survive to ages $\sim$10 Myr because their X-ray luminosities --- and, hence, disk photoevaporation rates --- are very low to begin with, and then further decline relatively early in their pre-MS evolution.  Such an interpretation of Fig.~\ref{fig:lxlbol} would be consistent with the suggestion that the well-established anticorrelation of $L_X$ and accretion rate in T Tauri stars \citep[e.g.,][and refs.\ therein]{Stelzer2012} is indicative of ``photoevaporation-starved T Tauri accretion'' \citep{Drake2009}. 

This simple interpretation is subject to several caveats. First, it requires that the values of $L_{X}$ (hence $\dot{M}_X$) presently measured for TWA M stars are, at least in relative terms, representative of these values over most  of the lifetime of the TWA. The comparisons with pre-MS M stars in Taurus (age $\stackrel{<}{\sim}$1 Myr) and IC 348 (age $\sim$3 Myr) presented in Figs.~\ref{fig:lxlbolsptype},~\ref{fig:lxlbol} and discussed in \S 2.3 would appear to support such an assertion.  

Second, because soft photons dominate the disk surface heating, Eq.\ 1 would yield inaccurate estimates of $\dot{M}_X$ for those stars whose X-ray spectral energy distributions (SEDs) differ significantly from the irradiating spectrum assumed in the \citet[][]{Owen2012} models. This model X-ray spectrum spans a range of plasma temperatures, peaking at $\log{T_X ({\rm K})} \approx7.3$. It is hence somewhat harder than the observed X-ray spectral distributions of TWA M stars (see \S 2.2 and Appendix B). If so, the predicted mass loss rates obtained via Eq.\ 1 would be systematically underestimated. On the other hand, the X-ray spectral fitting results obtained for the three TWA M star binaries analyzed here (see Appendix B) hint at the likelihood that $T_X$ is correlated with stellar mass \citep[see also][]{Johnstone2015}, and that the plasma emission model adopted by \citet[][]{Owen2012} more closely resembles the plasma parameters that are characteristic of early-M stars. If so, then Eq.\ 1 would be most directly applicable to these (higher-mass) M stars, but would tend to underestimate $\dot{M}_X$ for mid- to late-M stars.

\begin{table*}[!t]
\footnotesize
\begin{center}
\caption{\sc Dusty TWA M stars: accretion rates and predicted X-ray-driven disk mass loss rates}
\label{tbl:accretion}
\vspace{.1in}
\begin{tabular}{l|ccc|c}
\hline
\hline
            & \multicolumn{3}{|c}{\sc Observed} & {\sc Predicted} \\
{\sc Object} & $W_{10}$(H$\alpha$)$^a$ & $\log{\dot{M_{acc}}}$$^b$  
           & $\log{L_X}$ & $\log{\dot{M_X}}$$^c$ \\ 
           & (km s$^{-1}$) & ($M_\odot$ yr$^{-1}$) & (erg s$^-1$)  & ($M_\odot$ yr$^{-1}$) \\
\hline
TWA 3A & $\sim$250 & $-10.3$, $-9.6$ & 29.0 & $-9.1$ \\ 
TWA 30B & $\sim$200 & $-11.0$  & $<27.5$ & $<-10.6$ \\ 
TWA 31 & 447 & $-8.6$ & $<27.5$ & $<-10.5$ \\
TWA 30A & $\sim$230 & $-10.7$ & 28.0 & $-10.0$ \\ 
TWA 27 & 170, 320 & $-11.9$, $-9.8$ & $<26.0$ & $<-12$ \\ 
TWA 28 & 194 & $-12.8$ & $<26.7$ & $<-10.4$ \\ 
\hline
\end{tabular}
\end{center}

\footnotesize
Notes:\\
a) (Range of) width of H$\alpha$ emission line at 10\% of peak,
obtained from \citet[][]{Muzerolle2000, Stelzer2007,Looper2010a,Looper2010,Shkolnik2011}. \\
b) (Range of) inferred mass accretion rates, obtained from
\citet[][]{Muzerolle2000,
  Stelzer2007,Herczeg2009,Shkolnik2011}. Accretion rates for TWA 30A, 30B, 31 obtained from $W_{10}$(H$\alpha$) using Eq.\ 1 in \citet{Natta2004}. \\
c) Photoevaporative disk mass loss rate predicted from Eq.\ 1 given
stellar $L_X$.
\end{table*}

 Third, comparison of the theoretically predicted disk mass loss rates and mass loss timescales for accretion vs. EUV- and X-ray-driven photoevaporation indicates that photoevaporation should only become the dominant disk gas dispersal mechanism very late in disk evolution, after the bulk of the initial disk mass has already been lost via accretion \citep[][and references therein]{Owen2012,Gorti2015}. As a preliminary evaluation of the relative importance of accretion vs.\ photoevaporation for the TWA stars that appear to retain residual gaseous disks, we compare in Table~\ref{tbl:accretion} mass accretion diagnostics (i.e., the 10\% width of H$\alpha$ emission) and inferred mass accretion rates ($\dot{M}_{acc}$) with predicted values of (or upper limits on) $\dot{M}_X$ as obtained from Eq.\ 1. Based on the predictions for $\dot{M}_X$, it seems that the mass loss rate due to photoevaporation may equal or perhaps exceed that due to accretion in the cases of TWA 3A and 30A. In contrast, the inferred accretion rate of TWA 31 is at least two orders of magnitude larger than its predicted X-ray-driven photoevaporative mass loss rate. The comparisons of $\dot{M}_{acc}$ with $\dot{M}_X$ for the remaining three stars, all of which are undetected in X-rays, are inconclusive. 

\section{Summary and Conclusions}

To investigate the potential connection between the intense X-ray emission from young, low-mass stars and the lifetimes of their circumstellar, planet-forming disks, we have compiled a list of the X-ray luminosities ($L_X$) of all presently known M-type members of the 8 Myr-old TW Hya Association (TWA). The values of $L_X$ are mainly drawn from available archival data and published values in the recent literature. We have obtained $L_X$ for the individual components of the binary star systems TWA 8, 9, and 13 via analysis of archival {\it Chandra} data (Appendix B). 

Although our study suffers from poor statistics for stars later than M3, we find a trend of decreasing $L_X/L_{bol}$ with decreasing $T_{eff}$ for TWA M stars wherein the earliest-type (M0--M2) stars cluster near $\log{(L_X/L_{bol})} \approx -3.0$ and then $\log{(L_X/L_{bol})}$ decreases, and its distribution broadens, for types M4 and later. These mid- to late-M TWA stars generally appear underluminous in X-rays relative to very young pre-main sequence stars of similar spectral type and luminosity, consistent with previous studies indicating that mean $L_X$ declines more rapidly for ultra-low-mass stars and brown dwarfs than for early-M stars \citep[e.g.,][]{Stelzer2012}. This apparent decline of $L_X/L_{bol}$ near a spectral type of M4, if real, may reflect either the suppression of coronal X-ray emission by accretion or a fundamental difference between the internal structures of ultra-low-mass pre-MS stars and earlier-type pre-MS M stars. 

Notably, the fraction of TWA stars with evidence for residual primordial disk material also sharply increases for  subtypes of M4 and later, i.e., near the stellar effective temperature where $L_X$ decreases.  Furthermore, most of the newly discovered TWA candidates that are of late-M and L-type also display evidence for disks \citep{Rodriguez2015}. The extant data for the TWA hence suggest that disk survival times may be longer for ultra-low mass stars and brown dwarfs than for higher-mass (early-type) M stars --- a result that would be consistent with studies of the Upper Sco region \citep[age $\sim$10 Myr;][and refs.\ therein]{Scholz2007,Luhman2012}. These observations have interesting implications for models of disk evolution, which generally predict that any dependence of disk lifetime on stellar mass should be very weak \citep[e.g.,][]{Gorti2009,Owen2012}.

The apparent anticorrelation between the X-ray luminosities of low-mass TWA stars and the longevities of their circumstellar disks (Figs.~\ref{fig:lxlbolsptype}, \ref{fig:lxlbol}) could be interpreted to indicate that the persistent large X-ray fluxes from early-type M stars in the Association have contributed to the rapid dispersal of their primordial disks. Conversely, disks orbiting the  lowest-mass pre-MS stars and pre-MS brown dwarfs in the TWA may have survived because their relatively small X-ray luminosities have resulted in overall low disk photoevaporation rates.  

In the specific cases of TWA 3A and 30A, we infer that the rate of mass loss due to X-ray-driven photoevaporation may exceed that due to accretion. The dominant role of photoevaporation in the dispersal of these disks, if confirmed, would be consistent with their advanced ages. However, the (X-ray-undetected) star TWA 31 appears to be accreting disk mass far more rapidly than it could be losing disk mass loss via photoevaporation.

X-ray observations of additional mid- to late-type M stars in the TWA \citep[e.g.,][]{Rodriguez2015} and M-type members of similarly nearby young stellar groups, combined with acquisition of spectroscopic diagnostics of the disk gas masses and mass accretion rates of these same stars, are required to verify and further investigate the apparent coincidence of a dropoff of $L_X/L_{bol}$ and increase in primordial disk fraction for ultra-low-mass stars and brown dwarfs at an age of $\sim$10 Myr.  High-quality X-ray spectra of selected stars spanning the full range of M spectral types would also better inform photoevaporation models, by constraining the hardness of the X-ray radiation that is incident on protoplanetary disks orbiting the lowest-mass stars. 

\acknowledgments{\it This research was supported by NASA Astrophysics Data Analysis Program grant NNX12AH37G, NASA Exoplanets program grant NNX16AB43G, and National Science Foundation grant AST-1108950 to RIT. DP acknowledges a CONICYT-FONDECYT award (grant 3150550) and support from the Millennium Science Initiative (Chilean Ministry of Economy; grant Nucleus RC 130007). The authors thank Hao Shi for preliminary analysis that contributed to this study, and the referee, Manuel G\"udel, for helpful comments. }

\newpage

\section*{Appendix A: Notes on Inclusion and Exclusion of Individual Stars}

\begin{description}
\item[TWA 6:] This star was classified as K7 by \citet{Webb1999} and \citet{Manara2013} and as M0 by \citet{Torres2006} and \citet{HerHill2014}. We include TWA 6 in the sample, though we note that it appears to be at  the K/M spectral type borderline.
\item[TWA 9:] As \citet{Weinberger2013} and \citet{PecautMamajek2013} have noted, the {\it Hipparcos} parallax of TWA 9 casts some doubt on its membership in the TWA. Based on other considerations (e.g., Li line strength, proper motion, kinematic distance, location relative to other TWA members), \citet{PecautMamajek2013} conclude that TWA 9 is indeed likely a member of the Association, and that the parallax measurement is spurious. Hence, we retain TWA 9B in our TWA M star sample. As the higher-mass component TWA 9A is a mid-K star \citep{Webb1999}, our consideration of this component is restricted to the X-ray spectral analysis presented in Appendix B.
\item[TWA 14, 15, 21:] \citet[][and refs.\ therein]{Schneider2012} conclude these M stars are unlikely to be TWA members, and do not include them in their WISE-based assessment of the presence or absence of IR excesses. However, \citet{Ducourant2014} find these systems are indeed likely TWA members, based on convergent point analysis, so we include them in our sample.  
\item[TWA 18: ] Several studies have concluded that this system is unlikely to be a TWA member \citep[][]{Torres2008,Schneider2012,Ducourant2014}, so we have excluded it from our sample.
\item[TWA 22:] \citet{Mamajek2005} finds a low probability that TWA 22 (M5) is a TWA member; \citet{Teixeira2009} concluded this star is a member of the $\beta$ Pic Moving Group (rather than the TWA); and \citet{Ducourant2014} find that TWA 22 fails their convergence point membership analysis. Although \citet{Schneider2012} list this system as a possible member of the TWA, we exclude TWA 22 from our analysis.
\item[TWA 31:] \citet{Schneider2012} list this system as a possible member of the TWA, while \citet{Ducourant2014} find that it fails their convergence point membership analysis. While this star could hence be unrelated to the Association (at 110 pc, it is one of the more distant candidates), it displays strong, broad H$\alpha$ emission indicative of ongoing accretion \citep{Shkolnik2011}, suggesting it is not merely a young field star that happens to lie in the direction of the TWA. We retain TWA 31 in our sample.
\end{description}

\newpage

\section*{Appendix B: Chandra X-ray Spectral Analysis for TWA 8, 9 and 13}

\begin{figure*}[t!]
\centering
\includegraphics[height=2.5in,angle=0]{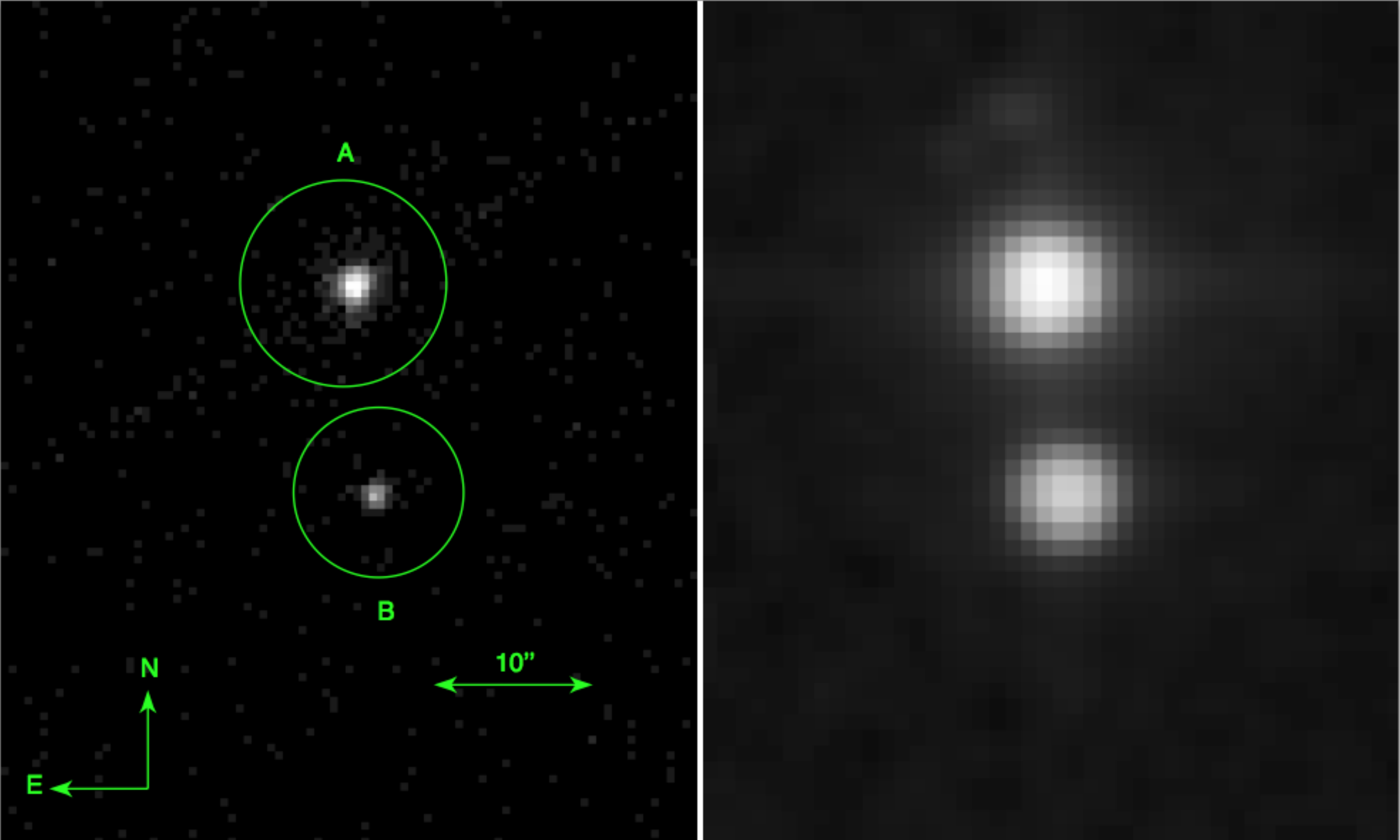}
\includegraphics[height=2.5in,angle=0]{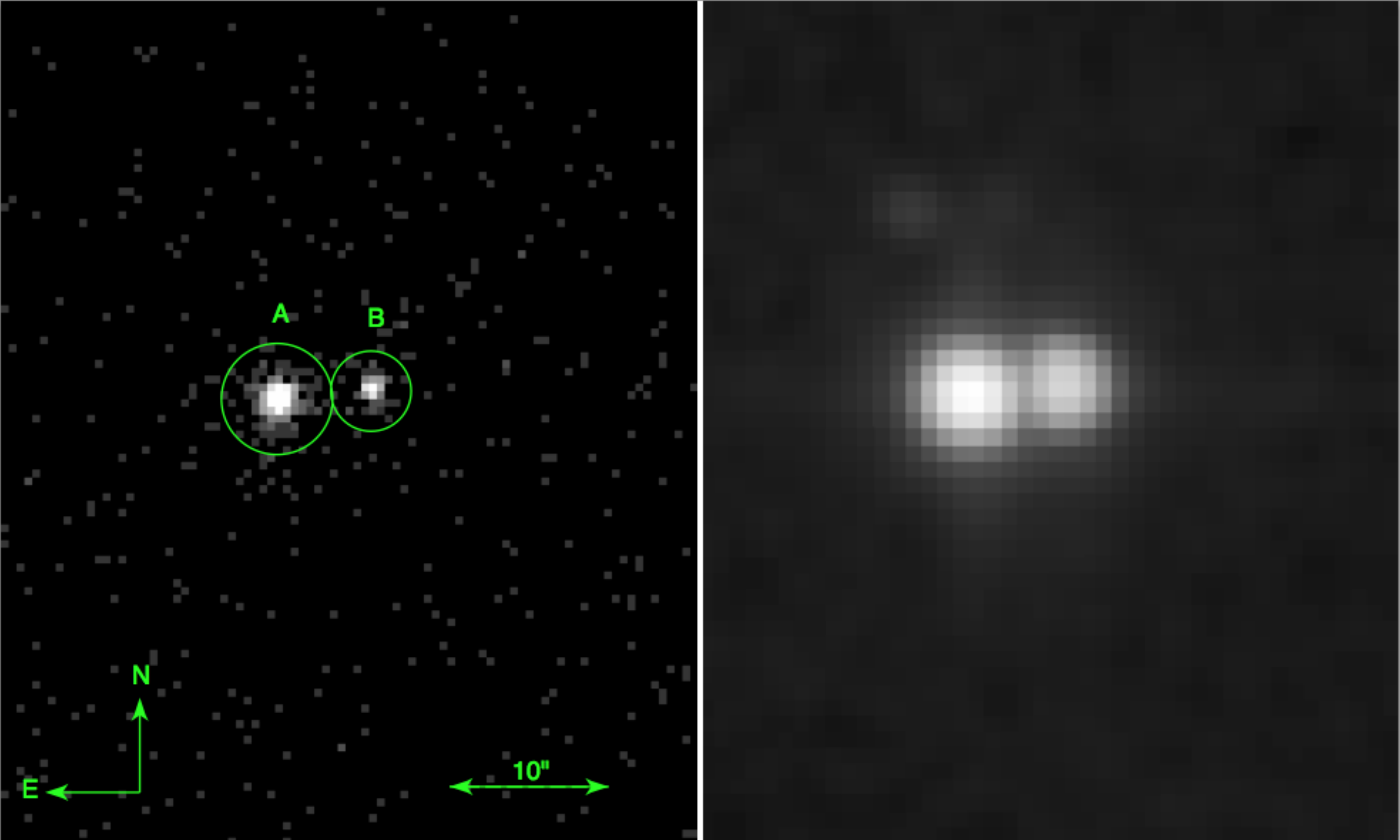}
\includegraphics[height=2.5in,angle=0]{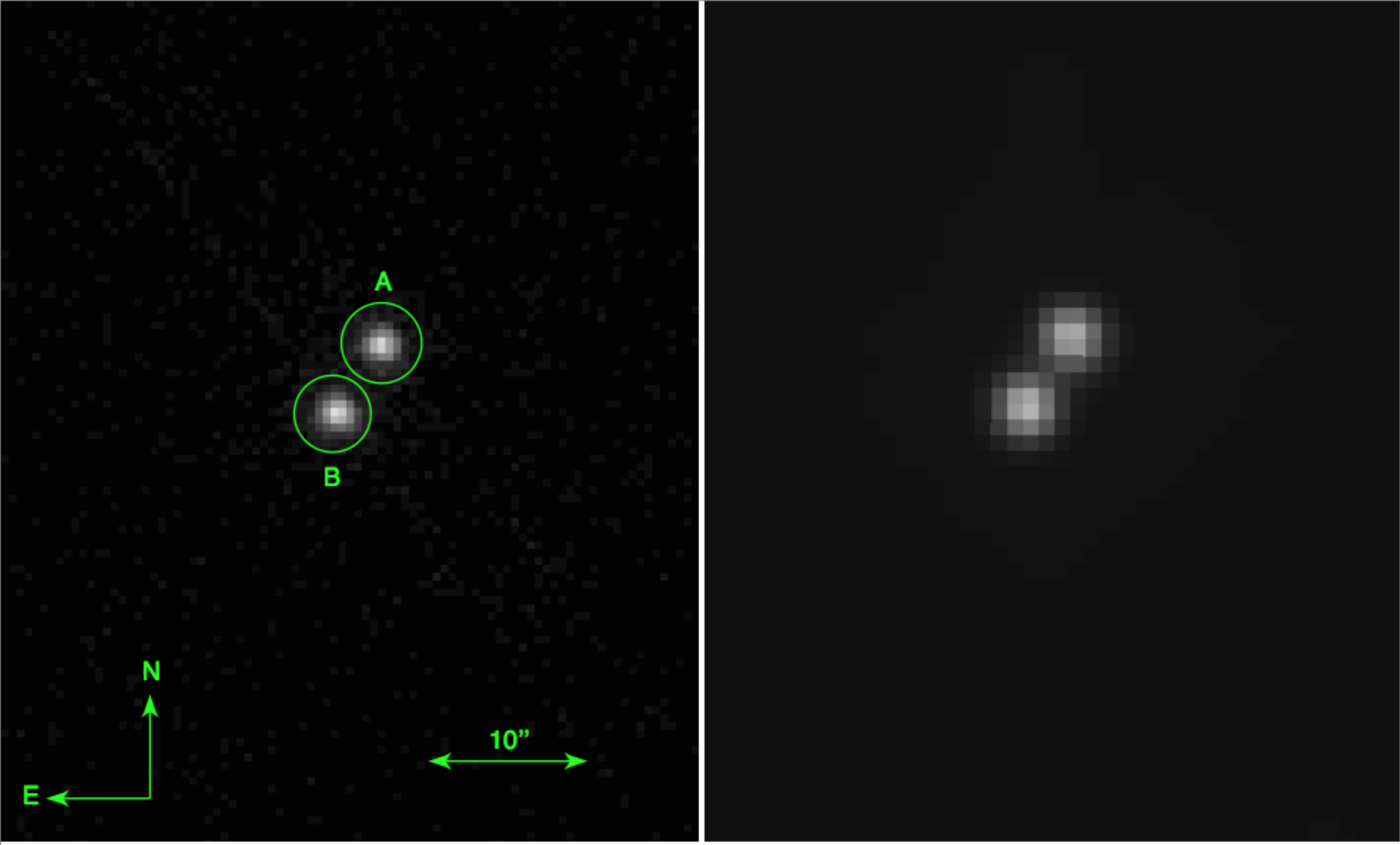}
\vspace{-.18in}
\caption{Chandra/ACIS-S3 images of TWA 8 (top), 9 (middle), and 13 (bottom), with orientation and scale indicated, alongside 2MASS J band images of the same systems at the same orientation and scale.}
\label{fig:TWAimages}
\end{figure*}

We have performed analysis of archival Chandra X-ray data for the M binary systems TWA 8, 9, and 13 \citep{Webb1999,Sterzik1999}. Five of the six components of these systems are M-type stars (Table 1), the lone exception being the K5 star TWA 9A. The binary separations of TWA 8, 9 and 13 are $\sim$13$''$, $6''$ and 5$''$, respectively. 
A summary of the available Chandra observations of the three systems is listed in Table~\ref{tbl:XRayObs}. All of these
data were obtained using CCD S3 of the Advanced CCD Imaging Spectrometer (ACIS) array. The resulting Chandra/ACIS-S3 images are displayed, alongside 2MASS J band images of the systems, in Fig.~\ref{fig:TWAimages}.

The pipeline-processed data files provided by the Chandra X-Ray Center were analyzed using standard science threads with CIAO version 4.7\footnote{http://asc.harvard.edu/ciao.}.
The CIAO processing used calibration data from CALDB version 4.6.5.
Spectra (and associated calibration data) were extracted within circular regions with diameters of $\sim$3--$8''$ centered on the stellar X-ray sources (see Fig.~\ref{fig:TWAimages}).
Background spectra were extracted within circular regions from nearby, source-free regions. The background-subtracted source count rates are listed in Table~\ref{tbl:XRayObs}.

Spectral fitting was performed with the HEASOFT \textit{Xanadu}\footnote{http://heasarc.gsfc.nasa.gov/docs/xanadu/xanadu.html.} software package (version 6.16) using XSPEC\footnote{http://heasarc.gsfc.nasa.gov/xanadu/xspec.} version 12.8.2.
We adopted the XSPEC optically thin thermal plasma model \verb+vapec+, whose parameters are the plasma abundances, temperature, and emission measure (via the model normalization). The potential effects of intervening absorption were included via XSPEC's \verb+wabs+ absorption model, although we found the absorbing column $N_H$ to be negligible in all cases (with the possible exception of TWA 13A, for which we find $N_H \sim 1.5\times10^{18}$ cm$^{-2}$).
We found that two temperature components are required to obtain acceptable fits to the spectra of four of the six stars, based on $\chi^{2}$ statistics (the exceptions being the relatively faint sources TWA 8B and 9B).
Initially, we fixed the parameters for the plasma abundances to values that have been determined to be typical of T Tauri stars in Taurus \citep[][and references therein]{Skinner2013}.
These abundance values (relative to solar) are H = 1.0, He = 1.0, C = 0.45, N = 0.79, O = 0.43, Ne = 0.83, Mg = 0.26, Al = 0.50, Si = 0.31, S = 0.42, Ar = 0.55, Ca = 0.195, Fe = 0.195, and Ni = 0.195.
We then allowed the abundances of Ne and Fe to be free parameters, since emission lines of these elements (plus O) tend to dominate the X-ray spectra of T Tauri stars in the $\sim$0.5--2.0 keV energy range \citep{Kastner2002}. Apart from the cases of TWA 8B and 9B, we find that leaving the Ne and Fe abundances free marginally improves the fit (slightly lowers $\chi^{2}$). In all cases, however, leaving Ne and Fe free only affects the results for X-ray flux (hence $L_X$) at the level of a few percent.

\begin{figure*}[t!]
\centering
\includegraphics[width=2.5in,angle=0]{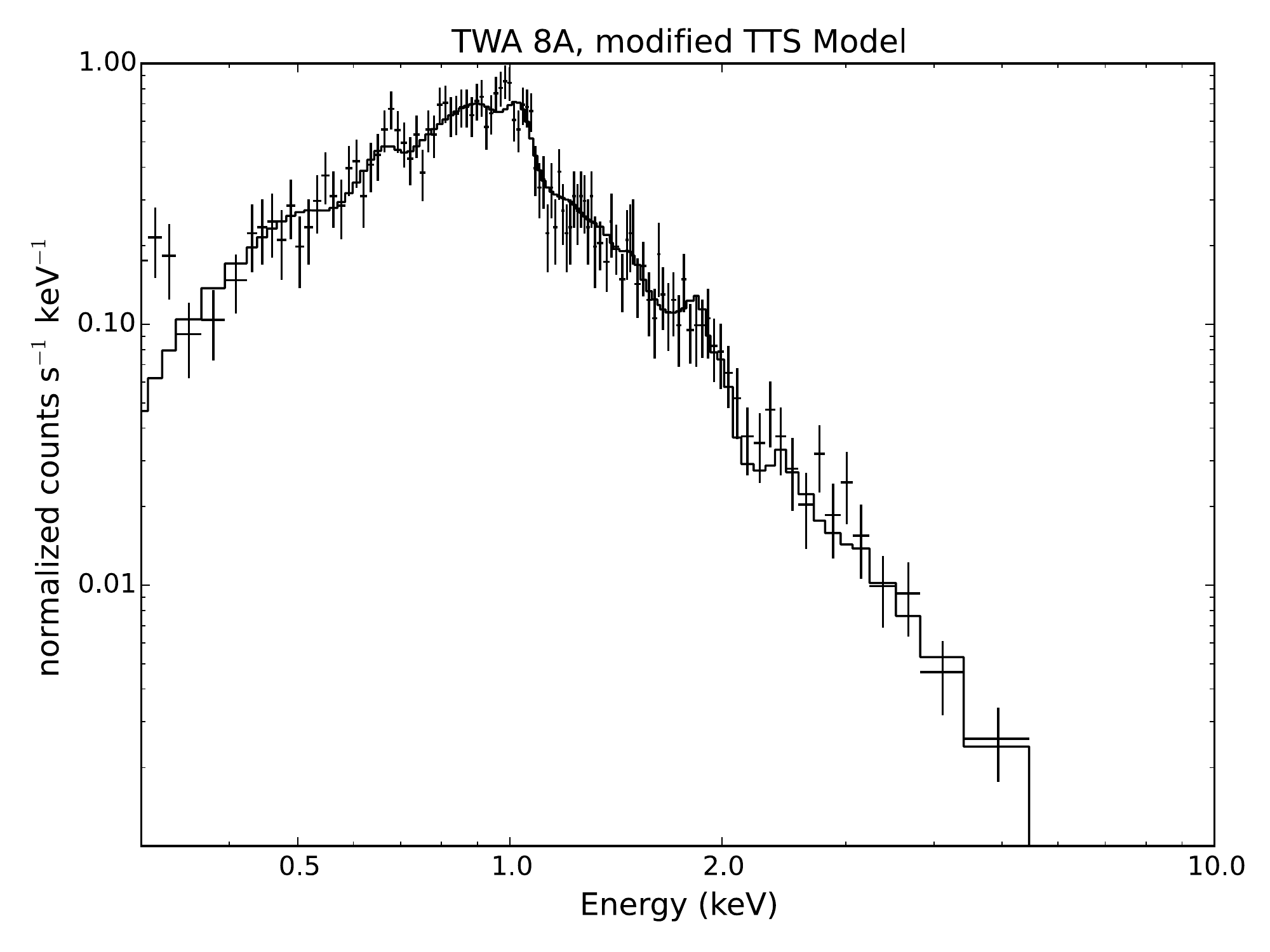}
\includegraphics[width=2.5in,angle=0]{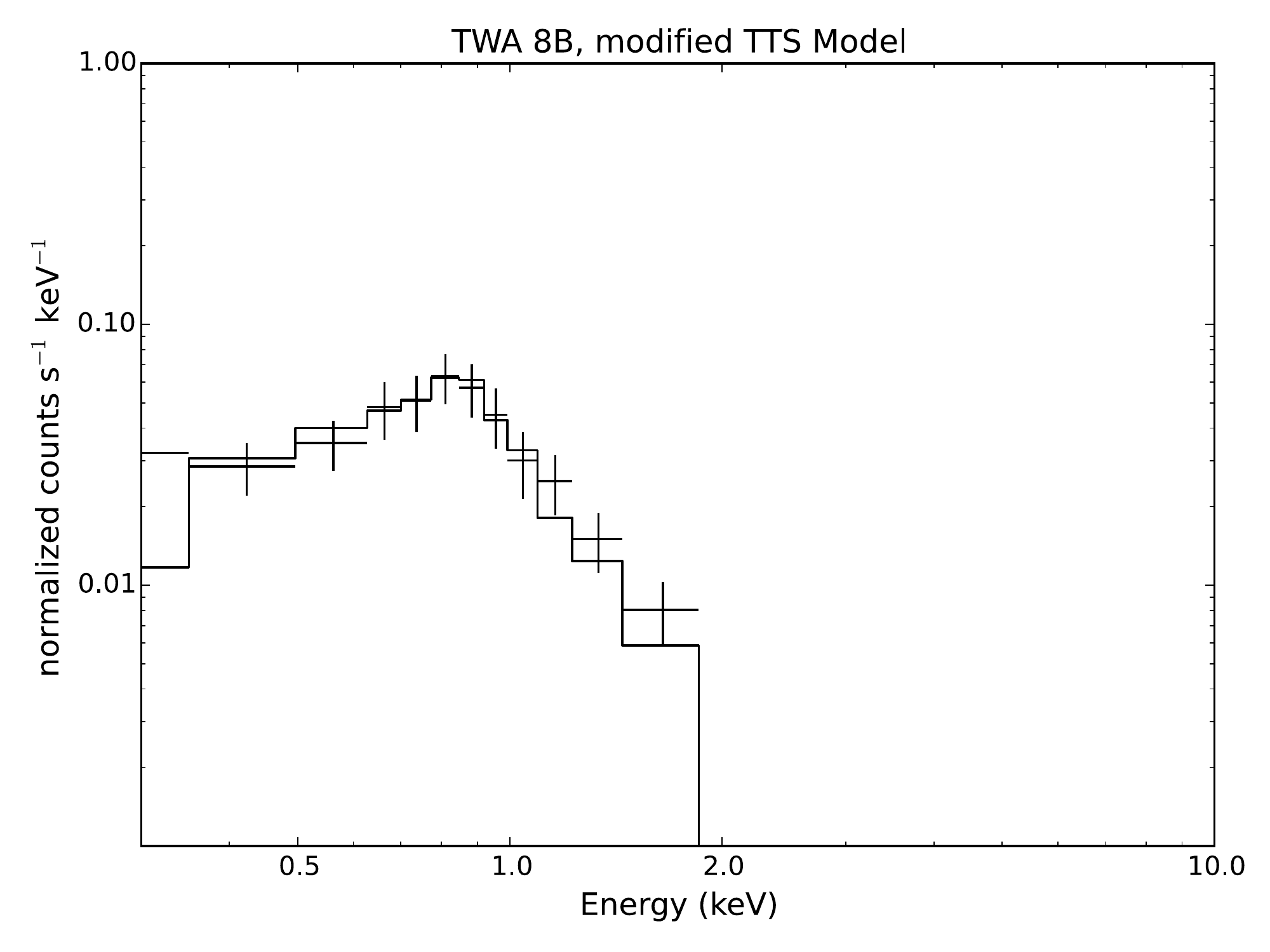}
\includegraphics[width=2.5in,angle=0]{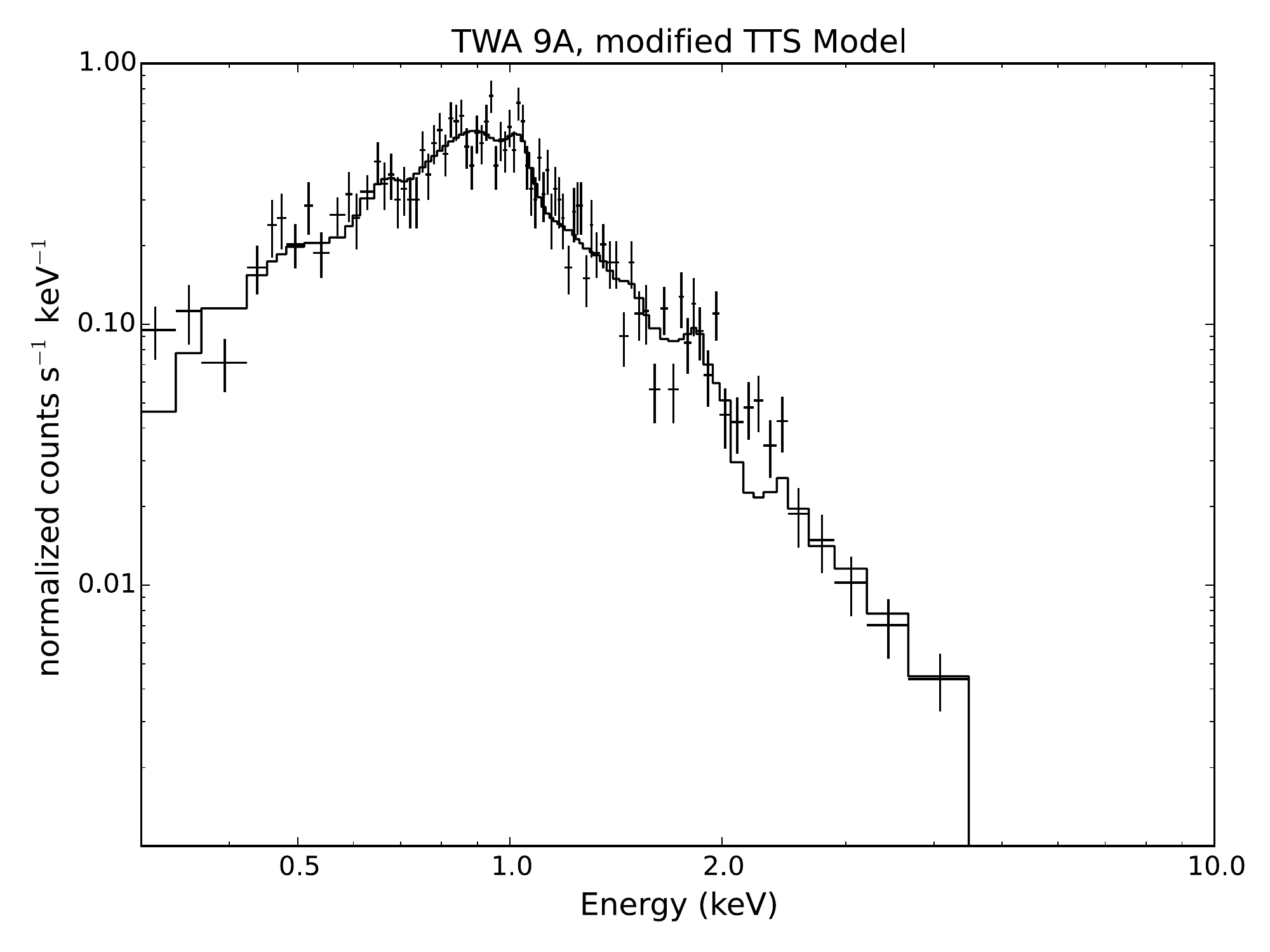}
\includegraphics[width=2.5in,angle=0]{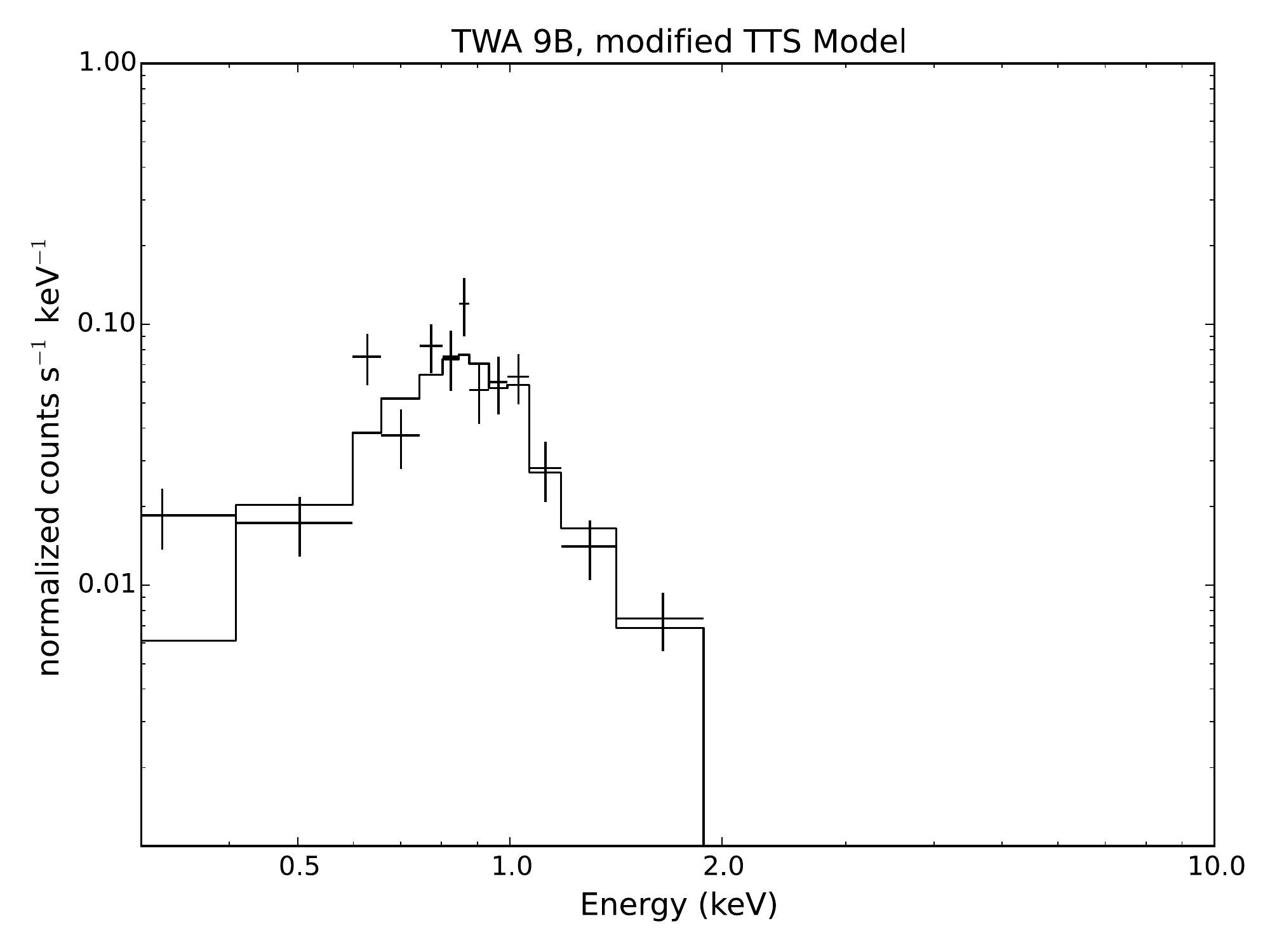}
\includegraphics[width=2.5in,angle=0]{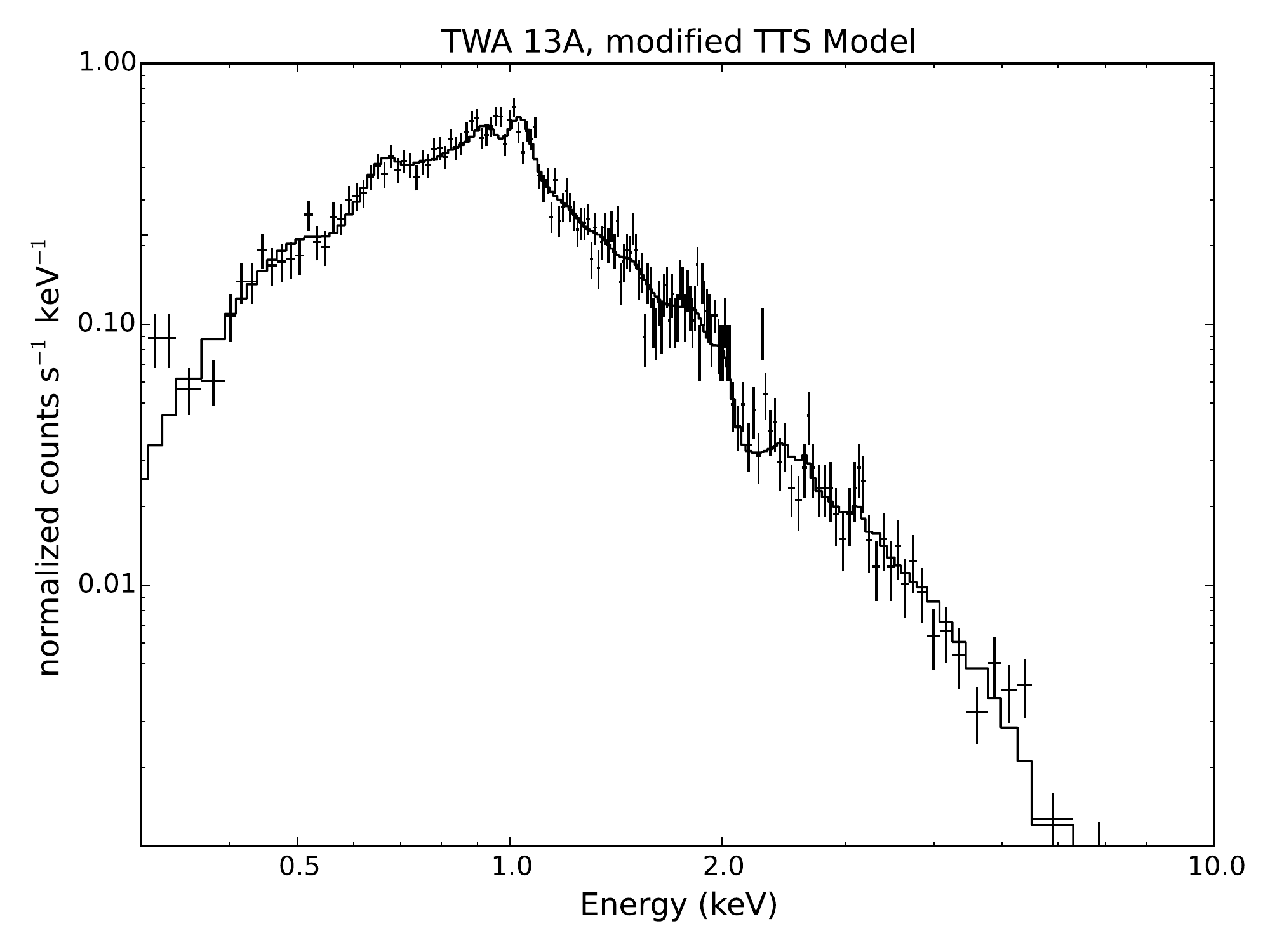}
\includegraphics[width=2.5in,angle=0]{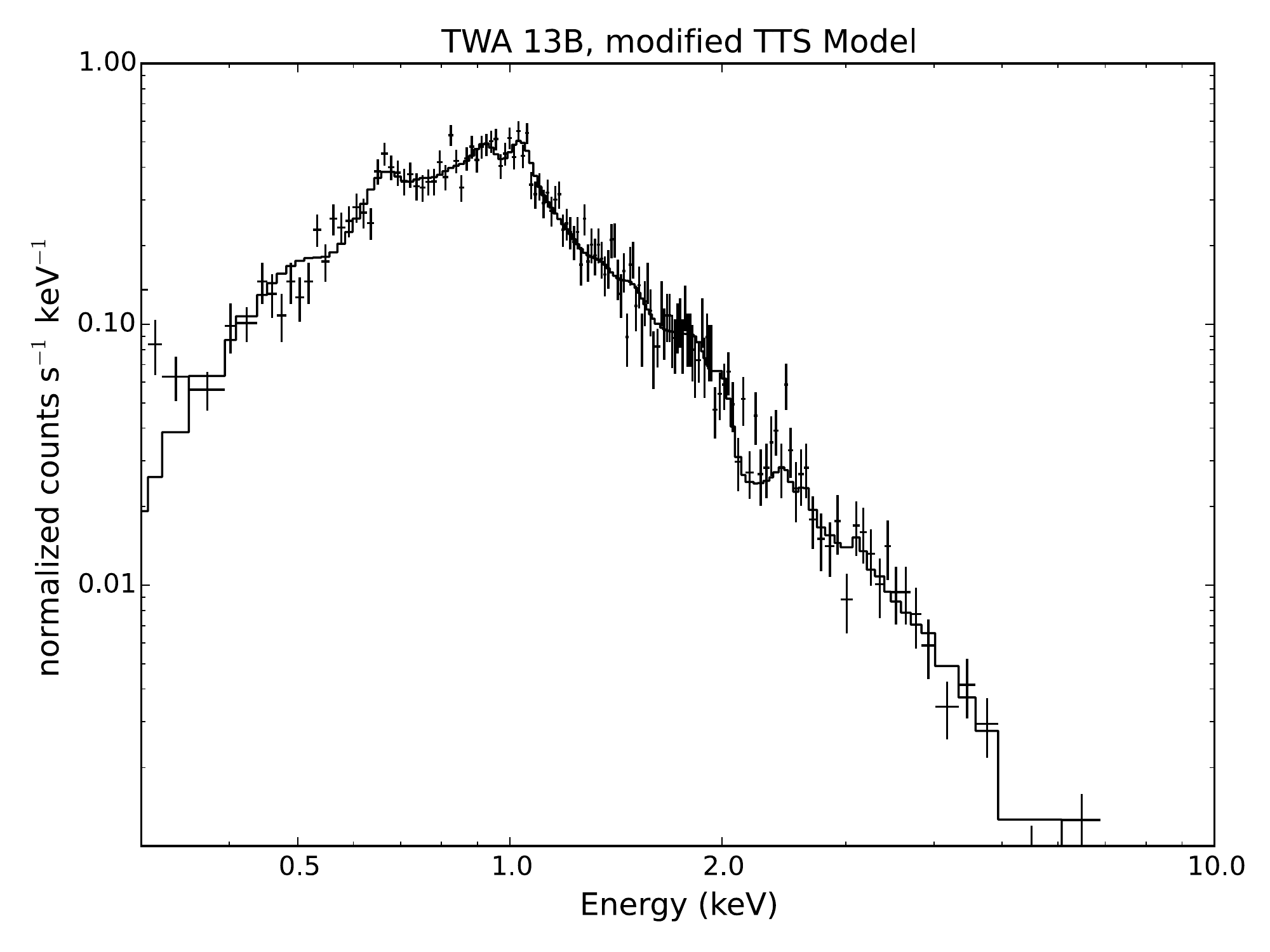}
\vspace{-.18in}
\caption{Chandra/ACIS X-ray spectra (crosses) of the A (left) and B (right) binary components of the TWA 8 (top row), 9 (middle row), and 13 (bottom row) systems, with best-fit two-component absorbed plasma models (histograms) overlaid. }
\label{fig:TWAfits}
\end{figure*}

The results of the spectral analysis are presented in Table~\ref{tbl:SpecAnalysis} and Fig.~\ref{fig:TWAfits}. The values for $L_X$ over the energy range 0.3--8.0 keV obtained from the fits are within $\sim$30\% of those determined by \citet{Brown2015} in all cases. We find that characteristic plasma temperature is correlated with $L_X$ for this small sample of TWA member stars. Specifically, the weakest X-ray sources (TWA 8B and 9B) have best-fit plasma temperatures $T_X\sim8$ MK, while the brighter sources (TWA 8A, 9A, 13A, and 13B) have characteristic (emission-measure-weighted) plasma temperatures $T_X\sim15$ MK. The results also indicate that TWA 8A may depart from the ``standard'' T Tauri star abundance pattern, in that it appears deficient in Ne and significantly overabundant in Fe. The other three stars for which we can extract plasma abundance results with some confidence --- TWA 9A, 13A, and 13B --- show marginal enhancements of Ne relative to the ``standard'' T Tauri star abundance.

\newpage



\begin{deluxetable}{c c c c}

\centering
\tablecolumns{4}
\tablewidth{400pt}
\tabletypesize{\footnotesize}

\tablecaption{\label{tbl:XRayObs}\sc \textit{Chandra} X-Ray
  Observations of TWA 8, 9, and 13}

\tablehead{

 & 
	\multicolumn{1}{c}{TWA 8} &
	\multicolumn{1}{c}{TWA 9} &
	\multicolumn{1}{c}{TWA 13}

}

\startdata
Obs.\ ID (PI) & \multicolumn{1}{c}{8569 (Herczeg)} & \multicolumn{1}{c}{8570 (Herczeg)} & \multicolumn{1}{c}{12389 (Brown)} \\
Exp.\ (ks) & \multicolumn{1}{c}{4.560} & \multicolumn{1}{c}{4.563} & \multicolumn{1}{c}{14.570} \\
Count Rate, Comp.\ A (ks$^{-1}$) & 615 (13) & 482 (10) & 449 (6) \\
Count Rate, Comp.\ B (ks$^{-1}$) & 47 (4)  & 52 (3)  & 543 (6)  \\

\enddata

\end{deluxetable}

\newpage
\clearpage

\begin{deluxetable}{c c c c c c c}

\centering
\tablecolumns{7}
\tablewidth{480pt}
\tabletypesize{\footnotesize}

\tablecaption{\label{tbl:SpecAnalysis}\sc X-Ray Spectral Analysis: Results}

\tablehead{

        \multicolumn{1}{c}{} &
	\multicolumn{6}{c}{Star}  \\

	\multicolumn{1}{c}{Parameter\tablenotemark{a}} &
	\multicolumn{1}{c}{TWA 8A} &
	\multicolumn{1}{c}{TWA 8B} &
	\multicolumn{1}{c}{TWA 9A} &
	\multicolumn{1}{c}{TWA 9B} &
	\multicolumn{1}{c}{TWA 13A} &
	\multicolumn{1}{c}{TWA 13B}
}
\startdata

Ne & 0.11 (0.03) & 0.4 (0.4) & 1.2 (0.4) & 1.1 (0.9) & 1.4 (0.2) & 1.6 (0.2) \\
Fe & 1.2 (0.3) & 0.11 (0.06) & 0.13 (0.04) & 0.12 (0.06) & 0.18 (0.04) & 0.15 (0.03) \\
$kT_{1}$ (keV) & 0.80 (0.06) & 0.68 (0.16) & 0.80 (0.05) & 0.68 (0.16) & 0.40 (0.03) & 0.44 (0.04) \\
$T_{1}$ (MK) & 9.3 & 7.9 & 9.3 & 7.9 & 4.7 & 5.1 \\
$kT_{2}$ (keV) & 2.5 (0.6) & ... & 2.5 (0.5) & ... & 2.1 (0.1) & 2.3 (0.1) \\
$T_{2}$ (MK) & 29 & ... & 29 & ... & 24 & 27 \\
$EM_{1}$ ($\times$$10^{52}$ cm$^{-3}$) & 3.44 & 0.36 & 3.47 & 0.26 & 2.46 & 2.69 \\
$EM_{2}$ ($\times$$10^{52}$ cm$^{-3}$) & 2.32 & ... & 2.90 & ... & 5.64 & 6.73 \\
$\chi_{red}^{2}$ & 0.78 & 2.97 & 1.41 & 2.72 & 1.36 & 1.49 \\
d.o.f. & 102 & 6 & 80 & 7 & 139 & 155 \\
$F_X$ ($\times10^{-12}$ ergs cm$^{-2}$ s$^{-1}$) & 2.8 & 0.26 & 2.2 & 0.20 & 2.5 & 3.1 \\
$L_{X}$ ($\times$10$^{29}$ ergs s$^{-1}$) & 6.3 & 0.46 & 7.2 & 0.64 & 9.2 & 11.1 \\
log $L_{X}/L_{bol}$ & $-$2.88 & $-$3.25 & ... & $-$3.38 & $-$2.93 & $-$2.85 \\

\enddata

\tablenotetext{a}{Fitting parameters are as follows: Ne, Fe are abundances of these elements
with respect to solar; $T_1$, $T_2$ and $EM_{1}$, $EM_{2}$ are, respectively, the
temperatures and emission measures of the two plasma components;
$\chi_{red}^{2}$ and ``d.o.f.'' are the reduced $\chi^2$ and degrees
of freedom of the fit, respectively; and $F_X$ and $L_X$ are the
intrinsic X-ray flux and X-ray luminosity, respectively.}
\end{deluxetable}


\end{document}